\documentclass{aastex63}

\usepackage{graphicx}
\usepackage{txfonts}

\graphicspath{{./}{figures/}}

\received{January 1, 2018}
\revised{January 7, 2018}
\accepted{\today}
\submitjournal{ApJ}

\shorttitle{Space Weather tool}
\shortauthors{Pagano et al.}

\begin{document}

\title{A New Space Weather Tool for Identifying Eruptive Active Regions}

\correspondingauthor{Paolo Pagano}
\email{pp25@st-andrews.ac.uk}

\author[0000-0001-5274-515X]{Paolo Pagano}
\affil{School of Mathematics and Statistics,
University of St Andrews, North Haugh,
St Andrews, KY16 9SS, Scotland}

\author[0000-0001-6065-8531]{Duncan H. Mackay}
\affiliation{School of Mathematics and Statistics,
University of St Andrews, North Haugh,
St Andrews, KY16 9SS, Scotland}

\author[0000-0003-2802-4381]{Stephanie L. Yardley}
\affiliation{School of Mathematics and Statistics,
University of St Andrews, North Haugh,
St Andrews, KY16 9SS, Scotland}



\begin{abstract}
One of the main goals of solar physics is the timely identification of eruptive active regions. Space missions such as Solar Orbiter or future Space Weather forecasting missions would largely benefit from this achievement. Our aim is to produce a relatively simple technique that can provide real time indications or predictions that an active region will produce an eruption. We expand on the theoretical work of \citet{Pagano2019fp} that was able to distinguish eruptive from non-eruptive active regions. From this we introduce a new operational metric that uses a combination of observed line-of-sight magnetograms, 3D data-driven simulations and the projection of the 3D simulations forward in time. Results show that the new metric correctly distinguishes active regions as eruptive when observable signatures of eruption have been identified and as non-eruptive when there are no observable signatures of eruption. After successfully distinguishing eruptive from non-eruptive active regions we illustrate how this metric may be used in a ``real-time" operational sense were three levels of warning are categorised. These categories are: high risk (red), medium risk (amber) and low risk (green) of eruption. Through considering individual cases we find that the separation into eruptive and non-eruptive active regions is more robust the longer the time series of observed magnetograms used to simulate the build up of magnetic stress and free magnetic energy within the  active region. Finally, we conclude that this proof of concept study delivers promising results where the ability to categorise the risk of an eruption is a major achievement.
\end{abstract}



\section{Introduction}
\label{introduction}

The ability to identify eruptive active regions prior to an eruption is crucial for improving
current space weather forecasting capabilities and to select targets for Solar Orbiter's 
observing campaigns. The fundamental aim is to predict as early as possible the location of the 
source region that the next solar eruption will originate from and correspondingly the time of 
the eruption.
Significant advances in this area have been made over the past few years due to the wealth of 
full disk data available from the Solar Dynamic Observatory (SDO). In particular, flare forecasting 
techniques have improved significantly.
For a summary of state-of-the-art flare forecast models we refer readers to \citet{2016ApJ...829...89B}.
To predict the occurrence of solar flares previous studies
have accomplished this by examining either the active region properties
from a physical perspective \citep{2007ApJ...661L.109G, 2017SoPh..292..159K,  2018SoPh..293...48J, 
2018SoPh..293...96K, 2018JSWSC...8A..34M}, or by using sophisticated computational techniques of 
data analysis
\citep{2013SoPh..283..157A, 2015ApJ...798..135B, 2015ApJ...812...51B, 2017ApJ...843..104L, 2017ApJ...835..156N, 2017ApJ...834...11R, 2018ApJ...858..113N, 2018SoPh..293...28F, 2018ApJ...853...90B, 2019ApJ...877..121L, 2019SoPh..294....6D}.
However, increasing our capability of identifying flaring regions alone is currently not enough to 
improve space weather forecasting, as flares are only one component of eruptive solar activity.
Space Weather events are largely produced by Coronal Mass Ejections (CMEs) that can be related 
to flares, but both phenomena can also occur without the other \citep{Gopalswamy2004}.
\citet{2018SSRv..214...46G} provides a thorough analysis of solar eruptions in relation to their predictability. A significant advancement in this area has been made by \citet{2018SoPh..293...60M}
which has combined two large-scale
projects to provide a crucial connection between the prediction of 
flares \citep[FLARECAST][]{2018cosp...42E1181G} and how this can support the prediction of CMEs 
studied in the HELCATS catalogue \citep{2018EGUGA..20.7441B}. Also, \citet{2017SpWea..15..577M} 
illustrated an example of how agencies are preparing for operations of flare forecasts. In addition to flares,
efforts have been deployed to predict the occurrence of CMEs, either using 
observable precursors \citep{2012SoPh..276..219B}, machine learning \citep{2016ApJ...821..127B} or
statistical approaches \citep{2018ApJS..236...15A}.

In parallel to these studies, a large number of papers have significantly improved our understanding of the physical mechanisms that cause solar eruptions. While predictive techniques tend to empirically identify key parameters that are proxies of the eruption process, other studies identify what physical mechanisms trigger CMEs.
\citet{2015SoPh..290.3457S} and \citet{2018SSRv..214...46G} provide recent overviews of the models of CMEs and flares,
while \citet{2015JApA...36..123P} considers aspects related to numerical simulations.
Very generally speaking, we can distinguish CME models in two main categories that
describe how the Lorentz force imbalance that triggers the eruption is generated.
We have MHD instability theories, where existing structures are in a metastable equilibrium, where the runaway from it can be catastrophic \citep{ForbesIsenberg1991}, or it can follow either the Kink \citep{Sakurai1976,TorokKliem2005}, or  Torus 
\citep{KliemTorok2006, 2010ApJ...718.1388D, Aulanier2010} instability.
Alternativelly, the force imbalance is generated by the removal of confining structures, such as in the Breakout model \citep{Antiochos1999,2008ApJ...683.1192L}.
Morever, other eruption mechanims can be triggered outside or at the border of active regions, 
such as blowout jets \citep{2017A&A...598A..41C}.

The aim of this paper is to provide a proof of concept study for a new approach to identify 
eruptive active regions based on the physical analysis provided by \citet{Pagano2019fp},
which we will now refer to as Paper I.
In Paper I, we analysed the 3D magnetic field evolution of a set of active regions
using the magnetofrictional relaxation model of \citet{Mackay2011}.
We then derived a theoretical metric
that discriminated eruptive from non-eruptive active regions.
We identified an empirical threshold (based on the small sample of active regions considered)
and we developed a magnetogram projection technique that allowed us to successfully 
apply the theoretical metric to both observed magnetograms and magnetograms projected forward in time.
Building on \citet{Pagano2019fp},
the goal of this paper is to develop a new operational metric
that can be quickly computed and used to determine
whether an active region is at risk of producing an eruption.
In order to achieve this goal, we first
determine the most suitable features in the time evolution of the theoretical metric that indicate the risk of an eruption and secondly we construct a method to summarise this analysis into a single value ranging from 0-1 that measures the risk of an active region producing an eruption.

The structure of the paper is as follows: in Sec.\ref{model} we summarise the data-driven 
magnetofrictional model used to simulate the 3D structure of the active regions, along with
the main properties of the active regions used in this study.
In Sec.\ref{predicttime} we discuss a new approach in the
application of projected magnetograms to predict the
future time evolution of the active regions. Next in Sec.\ref{lambdasection} we present the 
practical application that classifies eruptive and non-eruptive active regions in the context
of an operational model. In Sect.\ref{conclusions} we draw conclusions and consider future 
advancements of the technique.

\section{Summary of the model and simulations}
\label{model}

To simulate the data-driven, 3D evolution of active regions we use the magnetofrictional 
relaxation model of \citet{Mackay2011}. This technique uses a 3D cartesian domain where
the initial coronal magnetic field is extrapolated from the first magnetogram in the time series evolution of the active region and is assumed to be potential.
Next, the time sequence  of observed line-of-sight magnetograms is applied as an evolving lower boundary condition 
where the evolution injects free magnetic energy and helicity into the coronal field. The coronal
field then responds by evolving through a sequence of Nonlinear Force-Free states. As the initial 
condition is a potential field, we find that a ramp-up phase corresponding to approximately 35 
magnetograms (2 days) is needed before the coronal magnetic field significantly departs from this initial configuration.
After this time, the description of the active region magnetic field follows from the evolution of the
applied line-of-sight magnetograms as it has lost its memory of the initial condition.
Full details of the applied technique can be found in the papers
of \cite{Mackay2011}, \cite{Gibb2014}, and \citet{Yardley2018a}.

From the time series of the 3D magnetic field configurations for each active region, we derived a theoretical metric $\zeta\left(x,y,t\right)$ that can be used to 
distinguish eruptive from non-eruptive
active regions,
where $x$ and $y$ are the horizontal directions on the solar surface and $t$ is time. 
Full details regarding the computation of $\zeta\left(x,y,t\right)$ can be found in Section 3 of
Paper I. However, in simple terms $\zeta\left(x,y,t\right)$ is the product of three different functions that are also normalised:
\begin{equation}
\zeta\left(x,y,t\right)=\omega\left(x,y,t\right)\mu\left(x,y,t\right)\sigma\left(x,y,t\right).
\label{zeta2dt}
\end{equation}
The terms composing $\zeta\left(x,y,t\right)$ are now described:
(i) $\omega\left(x,y,t\right)$ is close to 1 at the location where a magnetic flux rope is present,
(ii) $\mu\left(x,y,t\right)$ is close to 1 where the Lorentz force is directed outwards and
(iii) $\sigma\left(x,y,t\right)$ is close to 1 where the Lorentz force is heterogeneous  in space.
In principle $\zeta\left(x,y,t\right)$ can take a value anywhere from 0-1 but in practise the value is found to be small with a strong spatial variation,
In Paper I, we studied eight individual active regions, five of which were associated 
with an observed eruption and three which were not.
Through considering the bi-dimensional time-dependent distribution of $\zeta\left(x,y,t\right)$ we 
extracted a time dependent quantity, $\zeta_{max}\left(t\right)$, i.e.
the maximum of $\zeta\left(x,y,t\right)$ at each time
that represents how the theoretical metric associated with an active region evolves in time.
Furthermore, a single number $\bar{\zeta}$, 
that is the time average of $\zeta_{max}\left(t\right)$ over an extended period of time (hours to $\sim1$ day),
is able to distinguish between eruptive and non-eruptive active regions.
Using our sample of active regions, we find an empirically derived threshold, $\bar{\zeta_{th}}=0.028$,
for the value of $\bar{\zeta}$ that discriminates between the eruptive and non-eruptive active regions.

An important aspect considered in Paper I is that this metric remains useful in discriminating eruptive and non-eruptive active regions when we introduce projected magnetograms.
Projected magnetograms were used to project the magnetic field evolution forward in time
using the most recently available observed magnetograms.
This was achieved by using the final two observed magnetograms to measure
the electric field at the lower boundary of the 3D magnetofrictional simulation,
as a function of space $(x,y)$ assuming it was constant thereafter, throughout the time of the projection.
In Paper I, we ran a set of  simulations where we replace observed magnetograms with projected ones and found that the metric $\bar{\zeta}$ is not significantly altered when enough observed magnetograms are used prior to the projection.

In the present paper, we develop a practical application of the theoretical approach presented in 
Paper I that illustrates the possible operational use of this technique. In this application, we test 
the possibility of using the data-driven magnetofrictional simulations of \citet{Mackay2011} combined with the
metrics identified in Paper I for the continuous monitoring of active regions.
We illustrate how the technique could be used to issue warnings
of the risk of eruptions when the observed and forecasted magnetic field configurations produce
values of $\bar{\zeta}$ above a critical value. To carry out this application we use a mixture of
observed magnetograms and projected magnetograms over a limited time period.
A key new aspect of the present paper, compared to the technique used in Paper I is that, (i) here we use projected magnetograms for a fixed time span only and (ii) we introduce a more advanced measure
to quantify the risk for an active region to produce an eruption.
This advanced metric builds in more information from the active region evolution 
compared to that used in Paper 1 and develops the theoretical metric into an operational context.
The work-flow of the technique is a follows:
\begin{itemize}
\item We first run a set of magnetofrictional simulations
where we vary the time when we switch from observed to projected magnetograms.
\item Next, we derive a function from the theoretical metric $\zeta\left(x,y,t\right)$ that
measures how the metric changes between the projected and observed evolution 
and how it relates to the empirical threshold identified in Paper I.
\item Based on this new metric we classify the state of the active region at a given time as 
either, red $\sim$ high risk, amber $\sim$ medium risk or green $\sim$ low 
risk to serve as a warning for potential forthcoming eruptions.
\end{itemize}
A key element of this study is the selection of active regions used to present the new technique.
Full details of the active regions can be found in Paper I and  \citet{Yardley2018a, Yardley2018b, Yardley2019} where Table\ref{activeregions} presents the main properties of the active regions. 
The technique applied in this paper is applicable to all types of active region. However, for the present study, we have focused on an initial small sample of young isolated active regions as this allows us to capture the full evolution of the active region (from birth to decay) in the NLFFF simulations. The magnetic field evolution of the active regions under consideration can then be analysed using the metric that is based only on the magnetic field properties and its time evolution.
In principle, any model that accurately reproduces the magnetic field configuration
and evolution of the active region can be used to apply this metric, even if some technical difficulties might need to be overcome.
Examples of potential technical challenges that need to be overcome in terms of the NLFFF model, such that arbitrary active regions can be studied are: i) maintaining a high spatial resolution over large fields
of view, ii) modelling active regions that have merged and undergone significant evolution and the injection of non-potentiality and iii) non-localised active regions where the equilibrium is perturbed by eruptions occurring elsewhere. In the future we will refine the NLFFF model used in the proof of concept study carried out here to tackle these challenges.
\begin{table}
\caption{Active region properties, as in \citet{Pagano2019fp}.
These active regions have been previously studied by
\citet{Rodkin2017}, \citet{James2018}, \citet{Yardley2018a}, \citet{Yardley2018b}, \citet{Yardley2019}.}
\label{activeregions}      
\centering                          
\begin{tabular}{c c c c c c}        
\hline\hline                 
 Active region & Observation Start & Observation End  & Magnetogram cadence & Eruption at & Publication \\    
\hline                        
AR11561 & 2012.08.29 19:12:05 & 2012.09.02 01:36:04 & 96 min & 2012.09.01 23:37 & Yardley+2018b, 2019 \\  
AR11680 & 2013.02.24 14:23:55 & 2013.03.03 19:11:56 & 96 min & 2013.03.03 17:27 & Yardley+2018b, 2019 \\
AR11437 & 2012.03.16 12:47:57 & 2012.03.21 01:35:58 & 96 min & 2012.03.20 14:46 & Yardley+2018a,b, 2019 \\
AR11261 & 2011.07.31 05:00:41 & 2011.08.02 06:00:41 & 60 min & 2011.08.02 05:54 & Rodkin+2017 \\
AR11504 & 2012.06.11 00:00:08 & 2012.06.14 22:24:08 & 96 min & 2012.06.14 13:52 & James+2018 \\
AR11480 & 2012.05.09 11:12:05 & 2012.05.14 00:00:05 & 96 min & none & Yardley+2018b, 2019 \\  
AR11813 & 2013.08.06 01:36:07 & 2013.08.12 00:00:07 & 96 min & none & Yardley+2018b, 2019 \\
AR12455 & 2015.11.13 04:47:55 & 2015.11.18 23:59:54 & 96 min & none & Yardley+2018b, 2019 \\

\hline                                   
\end{tabular}
\end{table}

\section{Rolling projection of the eruption metric}
\label{predicttime}

In Paper I, we have shown that through studying the evolution of the  3D configuration 
of the magnetic field of an active region it is possible to distinguish eruptive from non-eruptive active regions. We also concluded that the introduced theoretical metric was robust in 
identifying eruptive active regions when we use projected magnetograms as part of the
time sequence instead of observed magnetograms. In this section, we address whether
the same modelling technique can be used with a rolling projection time corresponding to 
10 magnetograms and how the metric derived from these simulations 
can be used to identify eruptive active regions. We note that the 10 magnetograms represents
either a projection time of $\sim10$ $hrs$ or $\sim16$ $hrs$ depending on whether the 
cadence of the magnetograms, $\Delta t$, is 60 or 96 minutes). 

To implement the rolling projection we define $t_0$ as the start time when we switch 
from observed to projected magnetograms. The value of $t_0$ must lie between the first 
and final observed magnetogram and, in practical terms, we vary $t_0$ such that in successive simulations 
more observed magnetograms are used prior to the period of projection. This reproduces the 
situation of observing an active region up to a given time, $t_0$ and then projecting its evolution 
forward in time.
From the time $t_0$ we then run the 3D magnetofrictional simulation forward for 
a time period equivalent to the acquisition of 10 
magnetograms, i.e. until the time $t=t_0+10 \Delta t$. The final time of the projected simulations
may be either before or after the time of the last observed magnetogram in our time series
given in Tab.\ref{activeregions}
depending on when the projection starts. For these projected simulations we 
compute $\bar{\zeta}$ by averaging the value of $\zeta_{max}\left(t\right)$ over only the 
time span of the 10 projected magnetograms.
This method of computing $\bar{\zeta}$ is different from 
that used in Paper I, as it is now only computed during the projected interval.
In our study we start from a value of $t_0$ that is at the end of the ramp-up phase
i.e. the initial phase during which the magnetic field loses memory of
its initial potential configuration.
To implement the rolling nature of the simulations after each period of projection we
return and increase $t_0$ by $\Delta t$ through using another observed magnetogram
before carrying out another projection. This process is repeated until
we reach the time $t_f$ of the final magnetogram in the observation time series.
In a real application, this study corresponds to starting from an initial 
time series of magnetogram observations and as soon as more observational information becomes available,
$t_0$ is increased by $\Delta t$, and the projection is 
repeated after the observed evolution is updated using the most recent measurements.
Such a technique, where our simulations 
temporal window moves can be used to determine whether the $\bar{\zeta}$ associated with 
an active region is going to be above or below the threshold for eruptions and whether it will 
increase or decrease. In an operational sense, this will partially address
the question of whether an eruption is becoming more or less likely.

\begin{figure}
\centering
\includegraphics[scale=0.28]{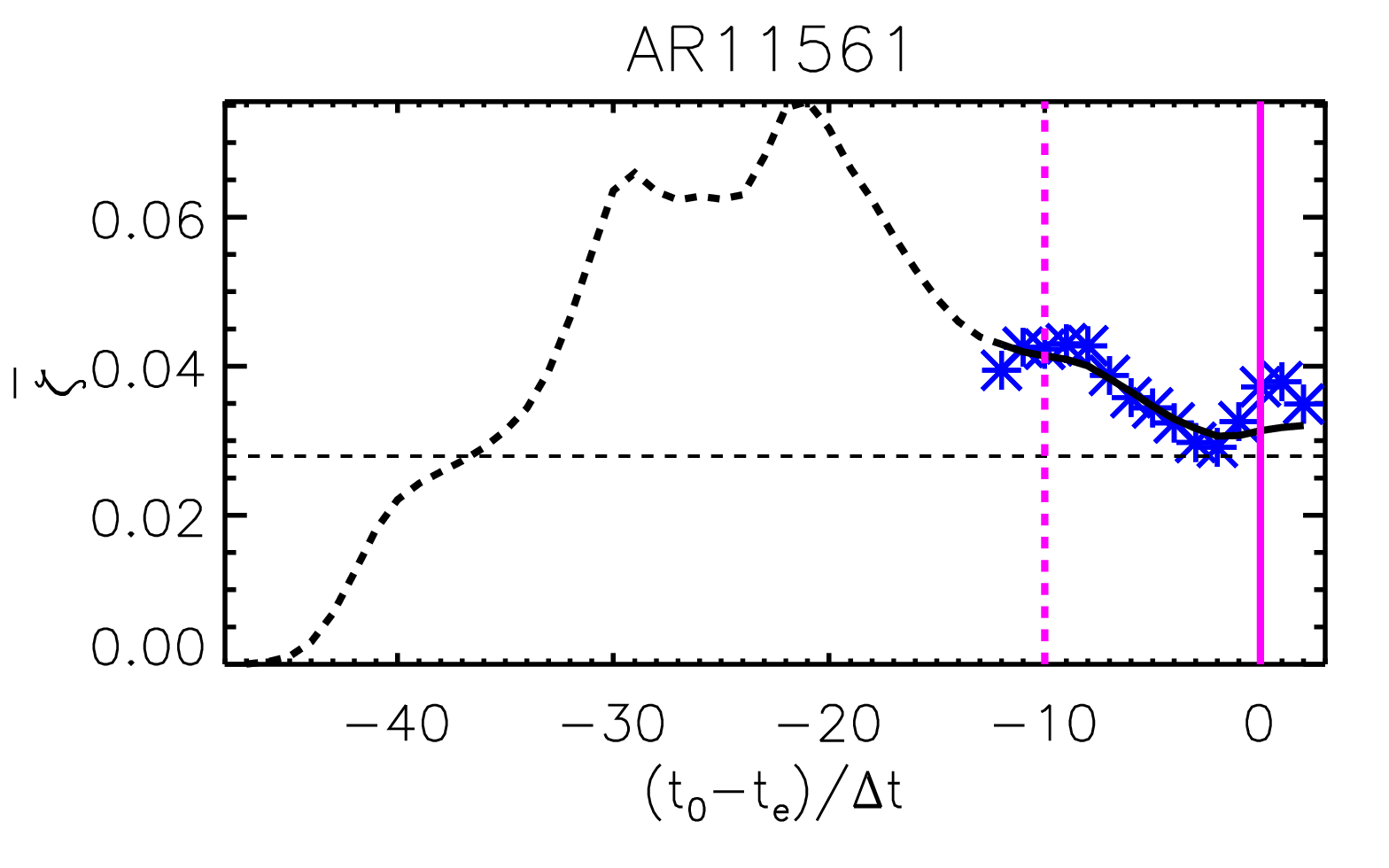}
\includegraphics[scale=0.28]{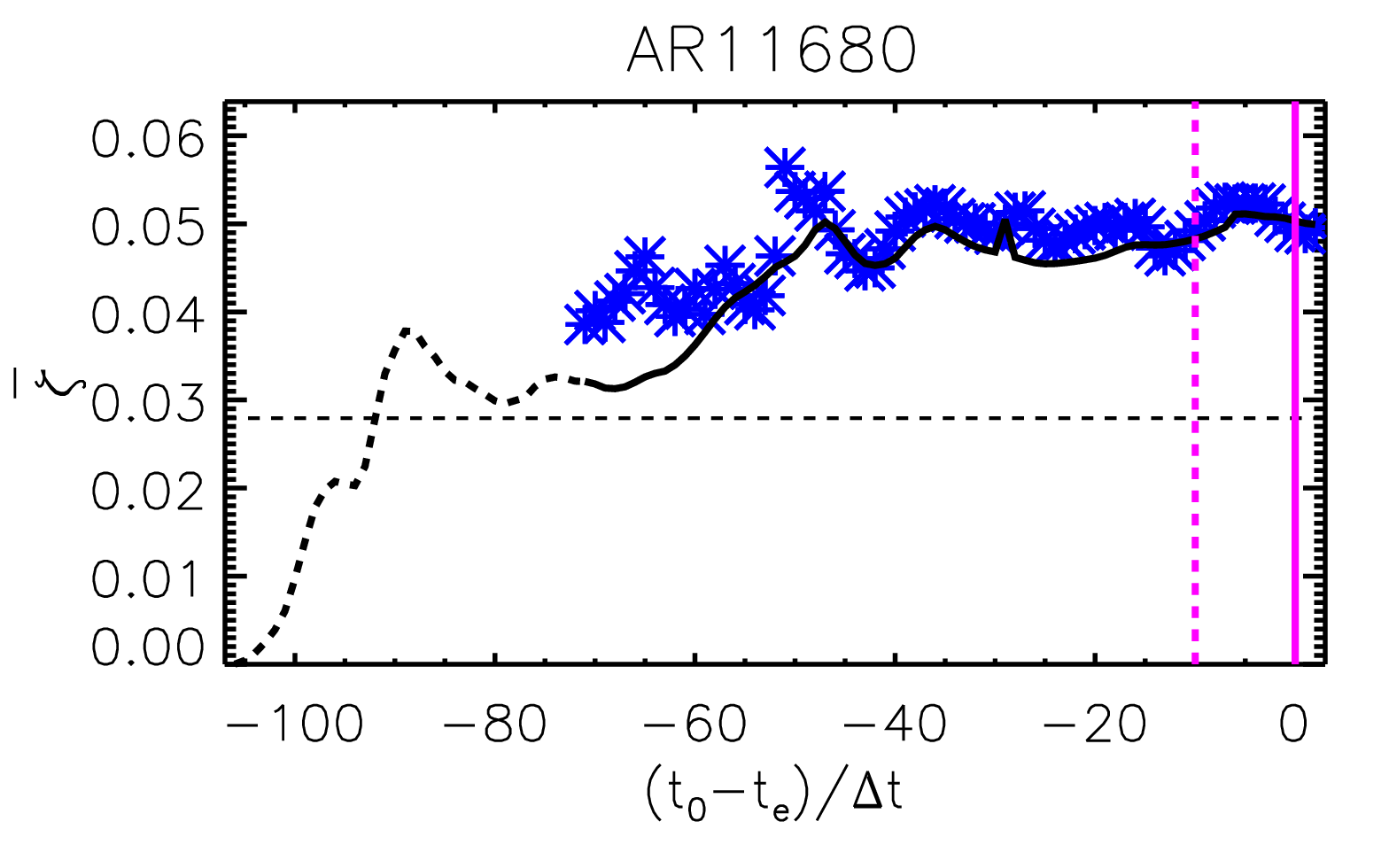}
\includegraphics[scale=0.28]{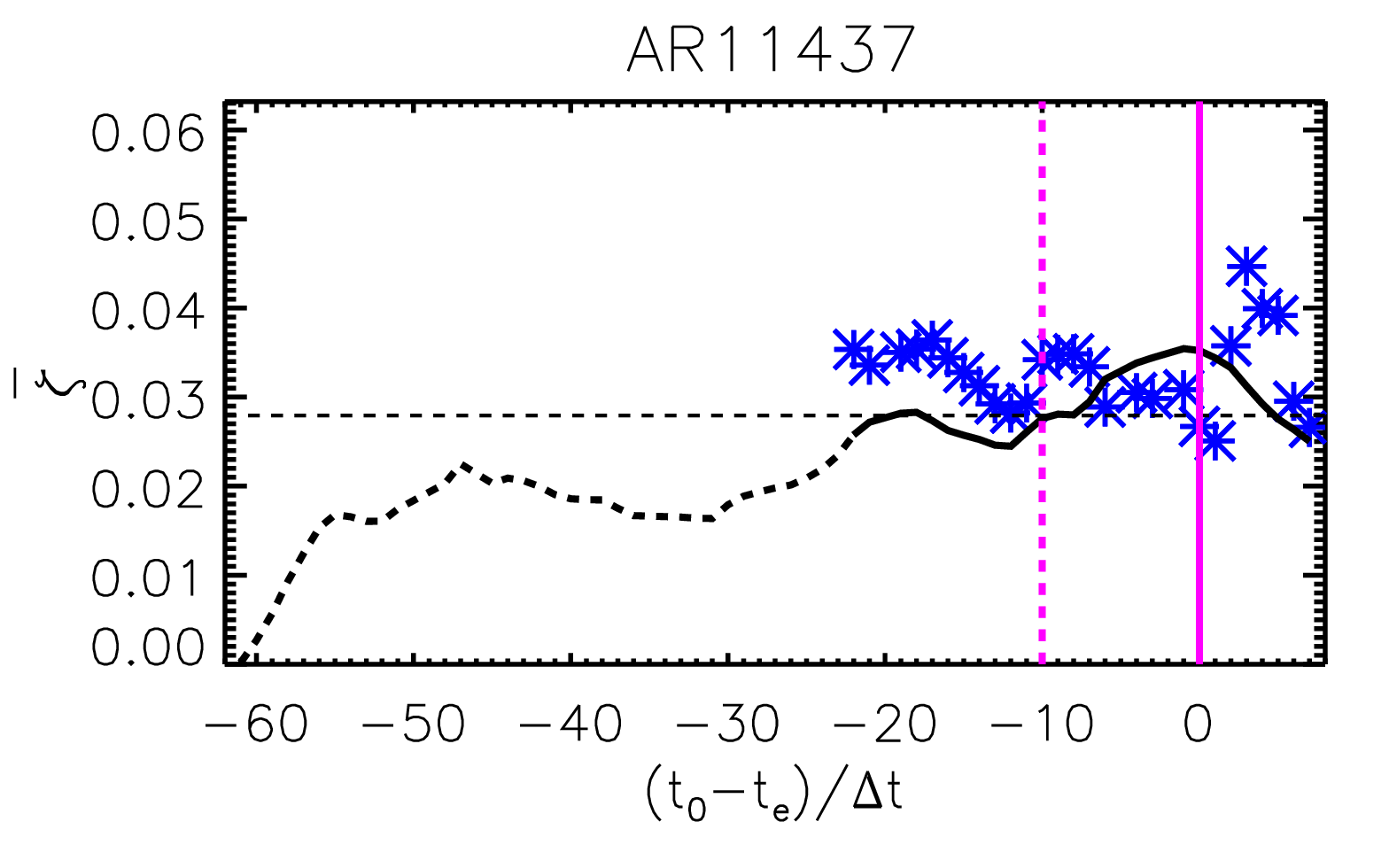}
\includegraphics[scale=0.28]{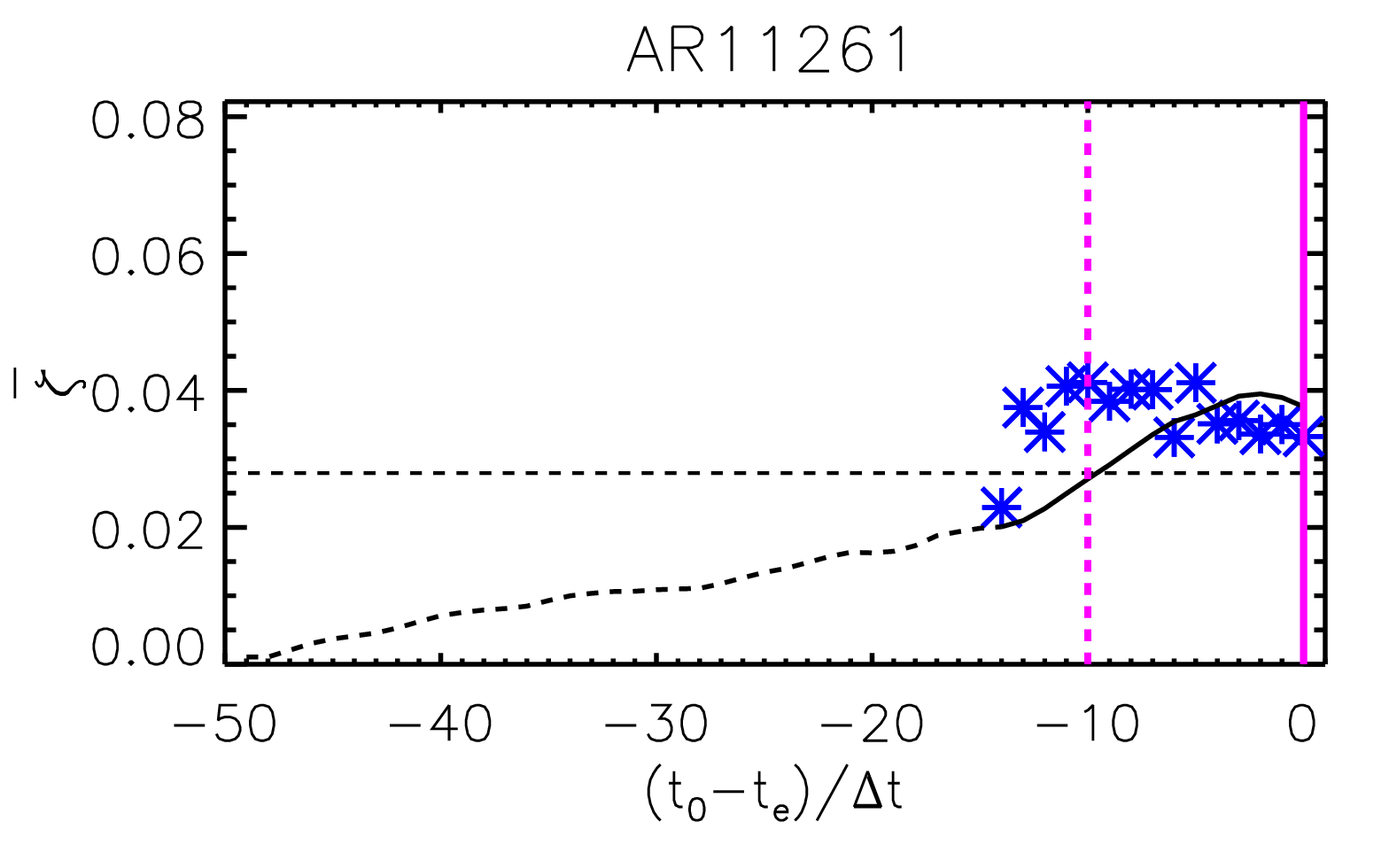}
\includegraphics[scale=0.28]{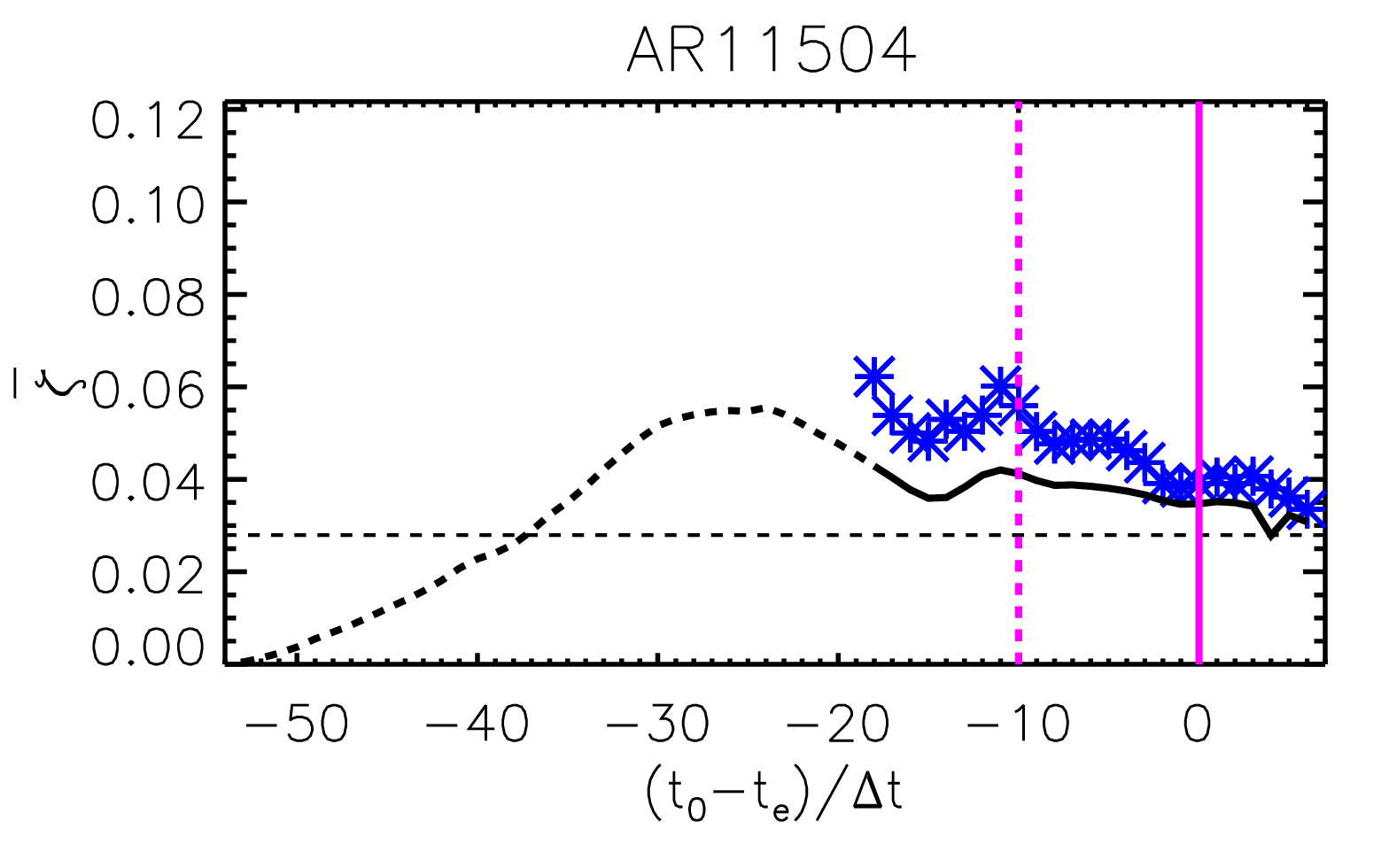}

\caption{The black line represents the evolution of $\zeta_{max}\left(t\right)$ for the eruptive active regions.
Blue asterisks at each $t_0-t_e/\Delta t$ are the values of $\bar{\zeta}$ obtained from the simulation that ran for 10 projected magnetograms after the observed one at $t_0$.
The magenta vertical lines represent the time of eruption, $t_e$ (continuous) and $t_0-t_e=-10\Delta t$ (dashed).
The horizontal dashed black line represents the value of $\bar{\zeta_{th}}=0.028$.}
\label{critprediction_eruptive_zetatime}
\end{figure}

Fig.\ref{critprediction_eruptive_zetatime} considers the value of $\bar{\zeta}$ obtained in 
these projected simulations (blue asterisks) as a function of $t_0$, shown for the eruptive active 
regions. In each plot the black curve gives the variation of the maximum of $\zeta\left(x,y,t\right)$,
deduced from Paper I (defined as $\zeta_{max}\left(t\right)$) when only observed magnetograms are used.
The black line is split into two segments where the black dashes are during the ramp-up phase of the 
magnetofrictional simulation and solid black line after the ramp-up phase.
In each plot the solid magenta line denotes the time of the observed eruption, $t_e$, where the plots 
are normalised with respect to $t_e$, such that $(t_0-t_e)/ \Delta t = 0$.
The dashed magenta line denotes the time, $(t_0-t_e)/\Delta t=-10$, such that the end time of the 
projected simulation occurs at the time of the eruption $t_e$.
In general the quantity $\bar{\zeta}$ determined for the projected magnetograms
is larger than the value of $\zeta_{max}\left(t\right)$ obtained from the full observational data set.
We now consider the evolution of $\bar{\zeta}$ and compare them with $\zeta_{max}\left(t\right)$ for each individual active region.

For AR11561 care must be taken, as we start using projected magnetograms 
immediately after the end of the ramp-up phase of the magnetofrictional simulation.
Due to this it has the least amount of prescribed observational evolution
prior to the start of the projections.
The value of $\bar{\zeta}$ at $(t_0-t_e)/\Delta t=-10$
is slightly larger than the value of $\zeta_{max}\left(t\right)$ at the same time.
Both quantities initially decrease in value, however they start to increase again 
shortly before the time of the observed eruption.
While all the values of $\bar{\zeta}$ remain above the indicated threshold, $\zeta_{th}$,
it is difficult here to identify a precise time when the active region
becomes eruptive from the projected simulation. However, the active region is classified as eruptive
throughout the entire projected time period.

As we already discussed in Paper I, AR11680 is an active region
that remains classified as eruptive for a significant period of time, 
as it consistently shows high values of $\zeta_{max}\left(t\right)$.
Over the time period of $-70 < (t_0-t_e)/\Delta t < -50$
the projected magnetogram simulations show an increasing value of $\bar{\zeta}$.
During this time period, $\zeta_{max}\left(t\right)$ also increases and
the values of $\bar{\zeta}$ and $\zeta_{max}\left(t\right)$ plateau together.
In this case, the only plausible conclusion is that an eruption is likely anytime
after $\bar{\zeta}$ plateaus.

For AR11437, just after the ramp-up phase there is a significant difference between $\bar{\zeta}$ and $\zeta_{max}\left(t\right)$.
The estimated value of $\bar{\zeta}$ then converges towards the value of $\zeta_{max}\left(t\right)$.
Near $(t_0-t_e)/\Delta t = -10$ 
the value of $\bar{\zeta}$ increases sharply to a value similar to that found
for $\zeta_{max}\left(t\right)$ at the eruption time $(t_0-t_e)/\Delta t = 0$. 
Then, the value of $\bar{\zeta}$ decreases after $(t_0-t_e)/\Delta t = -10$,
in the same way as $\zeta_{max}\left(t\right)$ does after the eruption time $(t_0-t_e)/\Delta t = 0$.
The fact that the evolution of $\bar{\zeta}$ about $(t_0-t_e)/\Delta t = -10$
resembles that found for $\zeta_{max}\left(t\right)$ about $(t_0-t_e)/\Delta t = 0$,
is a remarkable result for the magnetogram projection technique.
Additionally, in Paper I a decline in $\zeta_{max}\left(t\right)$ was explained 
by the occurrence of an eruption, after which the magnetic field releases free energy and evolves to a simpler configuration.
This simpler configuration is naturally associated with smaller values of $\zeta\left(x,y,t\right)$.

For AR11261 the magnetogram time series is probably too short to
produce conclusive results.
At $(t_0-t_e)/\Delta t = -10$ we find values of $\bar{\zeta}$
that are much larger than those found for $\zeta_{max}\left(t\right)$.
This can be attributed to one of two reasons. It could be related
to the occurrence of the eruption that follows shortly afterwards or it
could be due to the overestimation in $\bar{\zeta}$ that tends to occur
just after the ramp-up phase. However, this active region would also be 
characterised as eruptive.

Finally, AR11504 shows a behaviour similar to AR11437 where
the values of $\bar{\zeta}$ are larger than $\zeta_{max}\left(t\right)$
after the ramp-up phase. This is followed
by an increase in the value of $\bar{\zeta}$ near
$(t_0-t_e)/\Delta t = -10$ which is a signature of the 
eruption. Afterwards, the value of $\bar{\zeta}$ converges towards 
the value of  $\zeta_{max}\left(t\right)$ where both steadily decrease.

\begin{figure}
\centering

\includegraphics[scale=0.28]{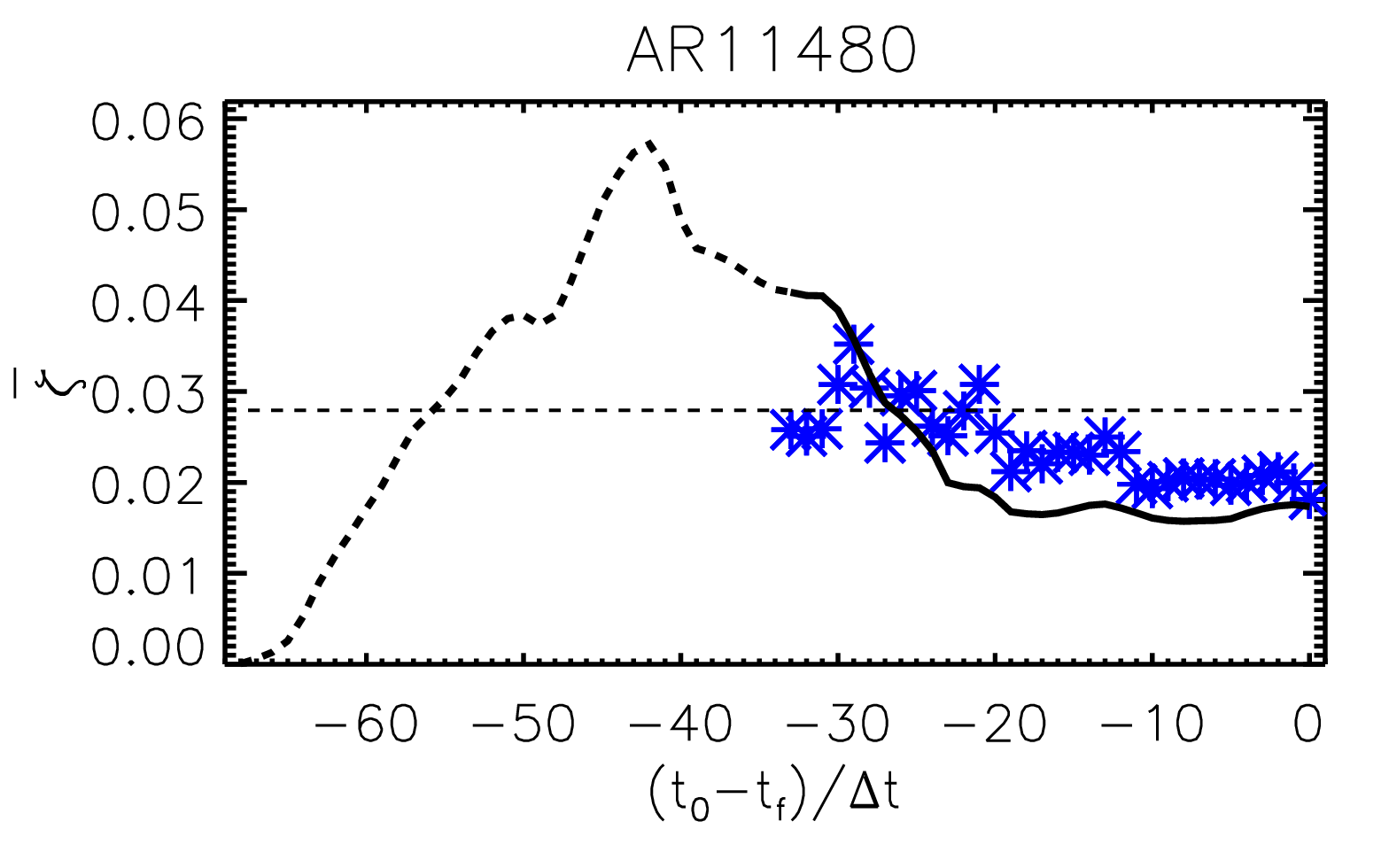}
\includegraphics[scale=0.28]{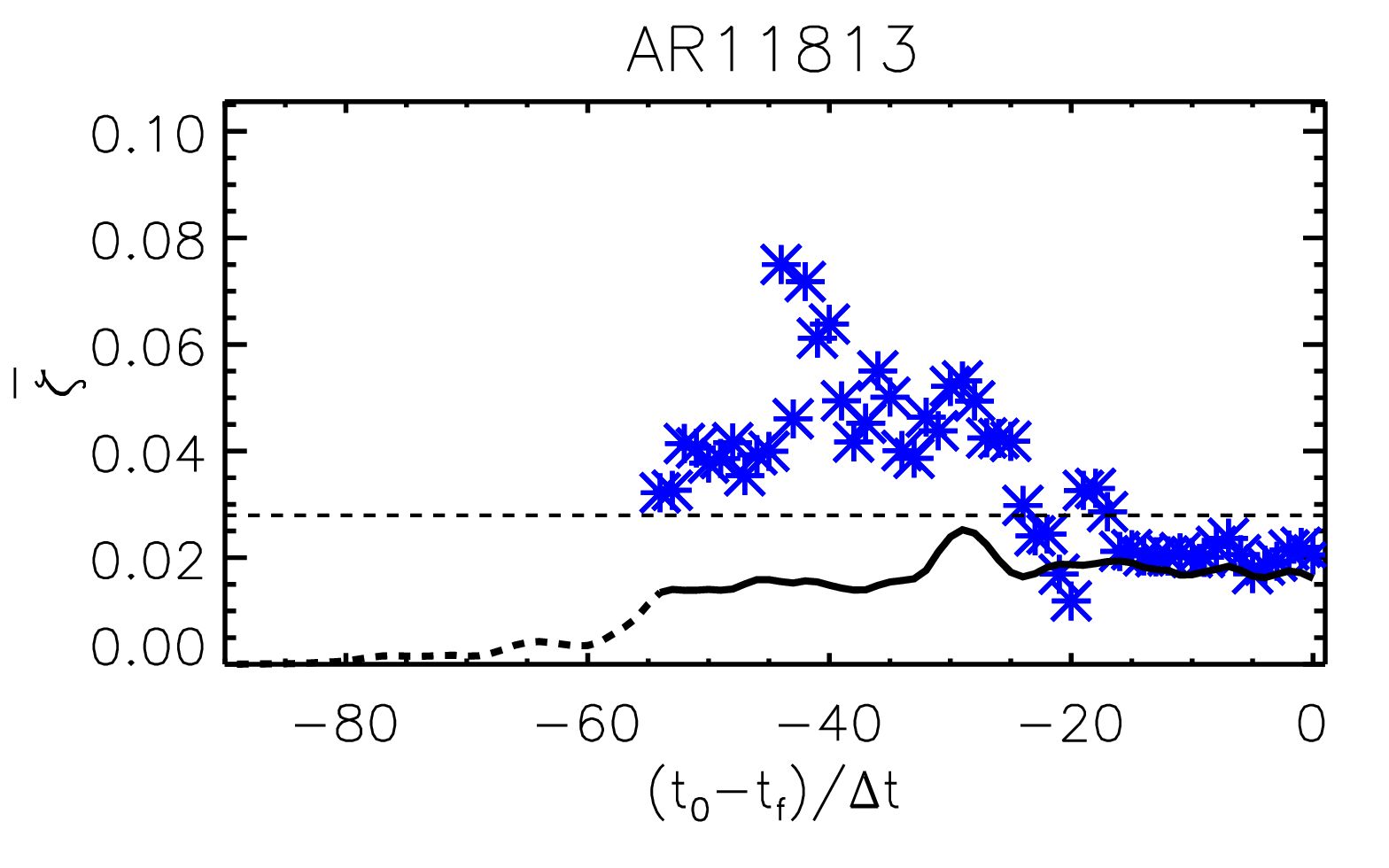}
\includegraphics[scale=0.28]{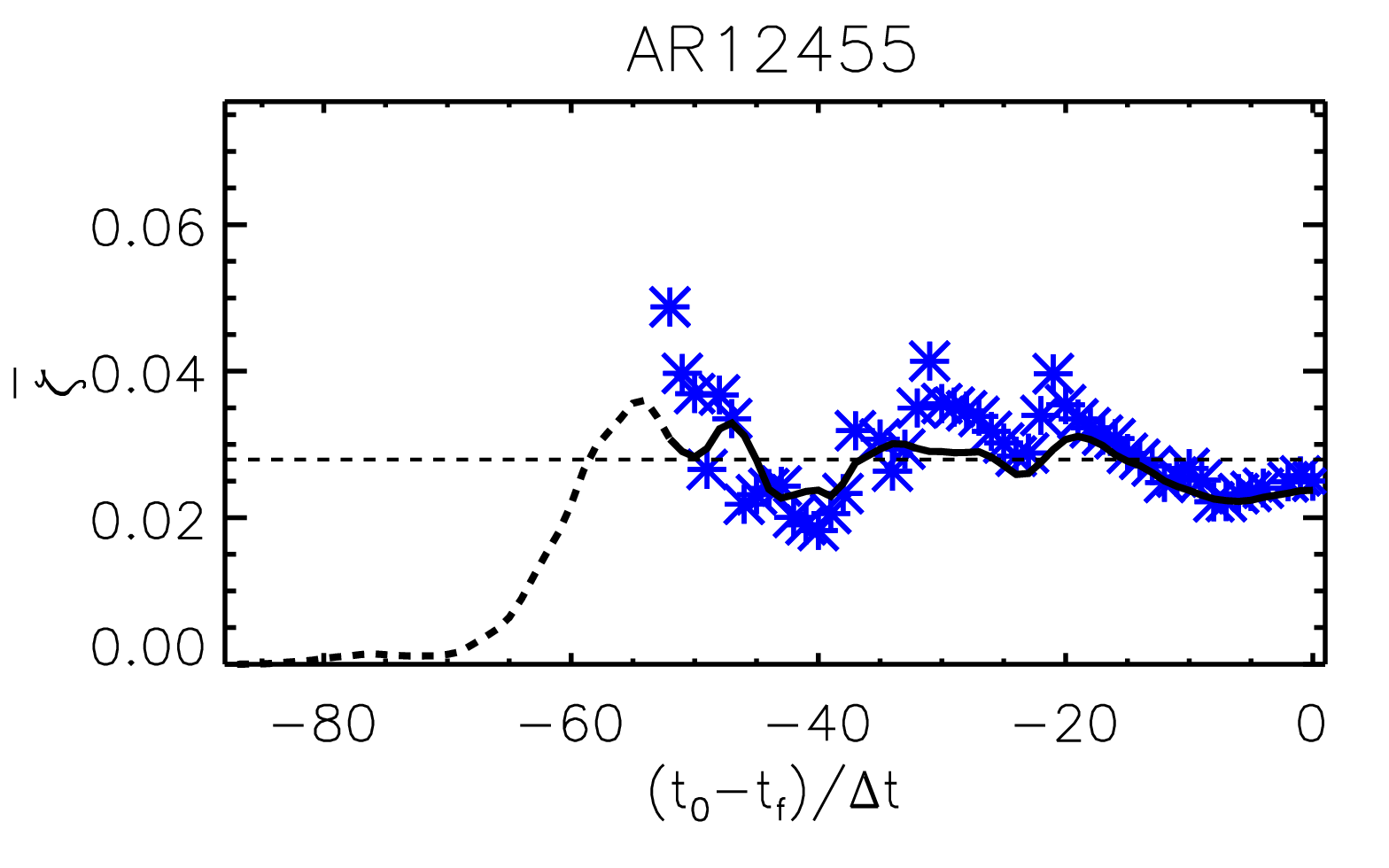}

\caption{The black line represents the evolution of $\zeta_{max}\left(t\right)$ for the non-eruptive active regions.
Blue asterisks at each $t_0-t_e/\Delta t$ are the values of $\bar{\zeta}$ obtained from the simulation that ran for 10 projected magnetograms past the observed one at $t_0$.
The horizontal dashed black line represents the value of $\bar{\zeta_{th}}=0.028$.}
\label{critprediction_noteruptive_zetatime}
\end{figure}

We also apply this technique to the non-eruptive active regions where the results
can be seen in Fig.\ref{critprediction_noteruptive_zetatime}. As these active
regions do not have any observed eruptions the results are presented
with respect to the time of the final observed magnetogram, $t_f$.
We find that for AR11480 the estimations of $\bar{\zeta}$
converge to a value under the threshold and therefore it would be 
classified as non-eruptive. For AR11813 the value of $\bar{\zeta}$ 
is initially very high after the ramp-up phase, but it converges to be below the threshold. Thus while during the early part of the projected evolution it would be classified as eruptive, for the later stages it would be non-eruptive. Finally, AR12455 presents a very oscillatory behaviour for both
$\zeta_{max}\left(t\right)$ and $\bar{\zeta}$, where both values oscillate above and below the threshold. The amplitude of the oscillations decay and near the end the values sit below the threshold.

All projected simulations show a value of $\bar{\zeta}$ that 
converges to a value that is larger/smaller than the empirical threshold 
$\bar{\zeta_{th}}$ (dashed line), for the eruptive/non-eruptive active regions
respectively. Thus the value of $\bar{\zeta}$ can discriminate between
eruptive and non-eruptive active regions over a prediction time period of
$\sim10$ magnetograms (10-16 hrs).

In summary, for eruptive active regions we find that the projected evolution of $\bar{\zeta}$ is consistently larger than the empirical threshold $\zeta_{th}$ that we have identified in Paper I.
This means that many of these active regions would be classified as eruptive.
While this is the case, the technique cannot currently provide an indication 
of the time of the eruption, and more work is needed in future to determine this.
For the non-eruptive active regions the projected values of $\bar{\zeta}$
can be highly scattered directly after the ramp-up phase,
e.g. when the projection technique overestimates the effect of flux emergence.
During this early stage the non-eruptive active regions may be initially classified as eruptive, but as the evolution proceeds,
the value of $\bar{\zeta}$ falls below the threshold and the active regions would be classified as non-eruptive. With the presently used theoretical metric the distinction
of eruptive and non-eruptive active region remains over a rolling projection time for this small sample study.
However, the values of the parameters found are subtle.
Therefore in the next section we aim to improve this metric 
into an operational framework by including more information such that we can more
clearly distinguish eruptive and non-eruptive active regions. 

\section{An Operational Metric for Predicting eruptive active regions.}
\label{lambdasection}

In this section, we develop a new method
based on the theoretical method introduced in Sec.\ref{predicttime}.
This new method will be more suitable to apply operationally.
The aim of this operational metric is to 
provide a measure of the eruptive condition of an active region 
at a given time $t_0$, where we run the magnetofrictional simulations
using observed magnetograms until $t_0$ and projected ones after $t_0$
until $t_0+10\Delta t$ (10 $\sim$ 16 hrs).
Our results from Paper I
(where we used only observed magnetograms or a long series of projected ones)
and from Sec.\ref{predicttime} (where we used a limited series of projected magnetograms)
clearly show that our theoretical metric $\bar{\zeta}$ discriminates between 
eruptive and non-eruptive active regions.
At the same time, an ad-hoc analysis was required to correctly interpret
the results of the metric from
the magnetofrictional simulations as the interpretation could be subtle.
For instance, we learned that i) we need to discard information obtained too close to the end 
of the ramp-up phase, ii) the theoretical metric naturally tends to decrease in the aftermath of an eruption and  iii) we need to compare the value of the metric with our empirical threshold in addition to its previous values.
To carry out this analysis manually in an operational context is not entirely realistic,
as we need the assessment to be carried out automatically by machines.
Therefore, there is the need for a new operational metric to be introduced that encompasses
the analysis given in Sect.\ref{predicttime} into one simple 
indicative number that measures the risk that an active region 
is going to produce an eruption or not.

The operational metric is defined by the function $\Lambda$ that depends on 
the past evolution, present configuration and projected evolution of the active region.
More specifically, it depends on values of 
$\bar{\zeta}$ when $\zeta_{max}\left(t\right)$ is averaged over different time windows, the instantaneous value of 
$\zeta_{max}\left(t\right)$, and the time derivative of $\zeta_{max}\left(t\right)$.
In the next section we outline how we derive the value of $\Lambda$.

\subsection{Computation of $\Lambda$}

To compute $\Lambda$ we use the following definitions:
\begin{itemize}
\item{$\zeta_{max}\left(t\right)$: the maximum value of $\zeta\left(x,y,t\right)$ at any given time t.}
\item{$\bar{\zeta}_{+10}$: the value of $\bar{\zeta}$ averaging $\zeta_{max}\left(t\right)$ over the time period $10\Delta t$ from $t=t_0$ (using only projected magnetograms).}
\item{$\bar{\zeta}_{-10}$: the value of $\bar{\zeta}$ averaging $\zeta_{max}\left(t\right)$ over the time period $10\Delta t$ until $t=t_0$ (using only observed magnetograms).}
\item{$\delta\zeta=\frac{\zeta_{max}\left(t\right)-\zeta_{max}\left(t-2\Delta t\right)}{2}$: the rate of change of $\zeta_{max}\left(t\right)$ between the magnetograms at $t=t_0$ and $t=t_0-2\Delta t$.}
\end{itemize}

Using these quantities we next define a set of arctan functions $\alpha$,
that can have values ranging from 0 to 1,

\begin{equation}
\alpha_{\zeta+10}=\frac{1}{\pi}\arctan\left(\frac{\bar{\zeta}_{+10}-\left(\bar{\zeta_{th}}+\beta_{th}\right)}{\beta_{th}}\right)+\frac{1}{2}
\label{alpha0}
\end{equation}

\begin{equation}
\alpha_{\zeta}=\frac{1}{\pi}\arctan\left(\frac{\zeta_{max}\left(t\right)-\left(\bar{\zeta_{th}}+\beta_{th}\right)}{\beta_{th}}\right)+\frac{1}{2}
\label{alpha1}
\end{equation}

\begin{equation}
\alpha_{+10-10}=\frac{1}{\pi}\arctan\left(\frac{\bar{\zeta}_{+10}-\bar{\zeta}_{-10}}{\beta_{+10-10}}\right)+\frac{1}{2}
\label{alpha2}
\end{equation}

\begin{equation}
\alpha_{n}=\frac{1}{\pi}\arctan\left(\frac{n-n_{ramp-up}}{0.5}\right)+\frac{1}{2}
\label{alpha3}
\end{equation}
where $n$ is the magnetogram index,  $n_{ramp-up}$ is the last magnetogram in the ramp-up phase,
$\beta_{th}$ and $\beta_{+10-10}$ are parameters that determine where and how steeply each $\alpha$ function transitions from values close to 0, to values close to 1.
In our application we use the following values for these parameters, $\beta_{th}=\beta_{+10-10}=10^{-4}$ and $n_{ramp-up}=35$. The latter value is chosen to be 35, as in Paper I we state that the 35th magnetogram of each sequence is the end of the ramp-up phase.

The rationale of the $\alpha$ functions is the following. 
Values of $\alpha_{\zeta+10}$ close to 1 lead to a higher eruption risk.
This happens when the value of $\bar{\zeta}_{+10}$ is larger than the threshold $\bar{\zeta_{th}}$,
which means that the active region is going to be in an eruptive state according
to the magnetic field evolution predicted by the projected magnetograms.
The same applies to $\alpha_{\zeta}$ that depends on how the instantaneous value of $\zeta_{max}\left(t\right)$ relates to the same threshold $\bar{\zeta_{th}}$.
Values of $\alpha_{+10-10}$ close to 1 also indicate an increased 
risk of eruption and are found when $\bar{\zeta}_{+10}$ is larger than $\bar{\zeta}_{-10}$.
This happens when the projected magnetograms predict that
the active region is going to be more likely to erupt, compared to what is measured
in the evolution of the prior 10 observed magnetograms.
Therefore $\alpha_{+10-10}$ measures the long term time variation of the metric over the time span corresponding to 20 magnetograms.
Finally, $\alpha_{n}$ takes values close to 1 when the time $t_0$ is far from the ramp-up phase of the magnetofrictional simulations.
This is to mitigate the impact of high values of $\bar{\zeta}_{-10}$,
due to the ramp-up phase giving false positives.

Finally, we also use another $\alpha$ function that gives negative values, between -1 and 0
\begin{equation}
\alpha_{\Delta\zeta}=\frac{1}{\pi}\arctan\left(\frac{\delta\zeta}{\beta_{\Delta\zeta}}\right)-\frac{1}{2}.
\label{alpha4}
\end{equation}
where $\beta_{\Delta\zeta}=10^{-3}$,
This function measures the short term time variation of the metric over the time span of 2 magnetograms
and has values close to -1 when the value of $\zeta_{max}\left(t\right)$ drops.
This helps identify times immediately after an eruption has occurred
where the risk of another one is reduced.

These parameter values ($\beta_{th}$, $\beta_{+10-10}$, and $\beta_{\Delta\zeta}$)
have been chosen such that the $\alpha$ functions switch 
from 0 to 1 within the range of values encountered for 
$\bar{\zeta}_{+10}$, $\bar{\zeta}_{-10}$, and $\zeta_{max}\left(t\right)$ in this present small sample study using young active regions.
Finally, we weight the $\alpha$ functions with
$w_{+10-10}=1$,
$w_{\zeta+10}=4$,
$w_{\zeta}=4$,
$w_{\Delta\zeta}=1$
and we compute $\Lambda$ as
\begin{equation}
\Lambda=\frac{w_{+10-10}\alpha_{+10-10}\alpha_{n}+w_{\zeta+10}\alpha_{\zeta+10}+w_{\zeta}\alpha_{\zeta}+w_{\Delta\zeta}\alpha_{\Delta\zeta}}{w_{+10-10}+w_{\zeta+10}+w_{\zeta}+w_{\Delta\zeta}}.
\label{lambda}
\end{equation}
It should be noted that only the function $\alpha_{+10-10}$ is multiplied by $\alpha_{n}$,
as it is the only $\alpha$ function that depends on values of the metric at times before $t_0$
and thus can be affected by the magnetic field description during the ramp-up phase.
In this work, the values for the weight parameters $w_{+10-10}$, $w_{\zeta+10}$, $w_{\zeta}$, $w_{\Delta\zeta}$ have been chosen
after a number of tests in order to best associate higher values of $\Lambda$ with active regions that are observed to erupt and,
more specifically, with the time $t_0$ falling within $t_e-10\Delta t<t_0< t_e$.
While the qualitative contribution of each $\alpha$ function
to $\Lambda$ is physically motivated and deduced from the properties of
the metric $\bar{\zeta}$ defined in Paper I,
the quantitative parameters used in each may still need to be refined. In this preliminary study we have selected by trial and error
a set of parameters and weights for the $\alpha$ functions that return high risk of eruption
for the eruptive active regions in the time period before the observed eruptions and low risk of eruption outside of this time period
or for non-eruptive active regions.
We note that it is possible to find unique values of parameters and weights that give consistent results for all the 8 active regions considered. However, care must be taken in the analysis of larger AR samples that contain a wider range of active regions as these values may depend on the stage of their evolution or vary with different types of active region. Finally, with this new operational metric defined by $\Lambda$ we expect active regions to fall in the full range of values from 0-1.

\subsection{Operational Application of $\Lambda$}

To be able to use the function $\Lambda$ for an operational application, we empirically
define two thresholds to quantify when an active region is entering an eruptive state.
We consider that an eruption is likely (eruptive active region) when $\Lambda \geq 0.55$, it is not likely (non-eruptive active region) when $\Lambda<0.35$, and 
the active region is borderline eruptive in the range $0.35<\Lambda\leq0.55$.

\begin{figure}
\centering
\includegraphics[scale=0.28]{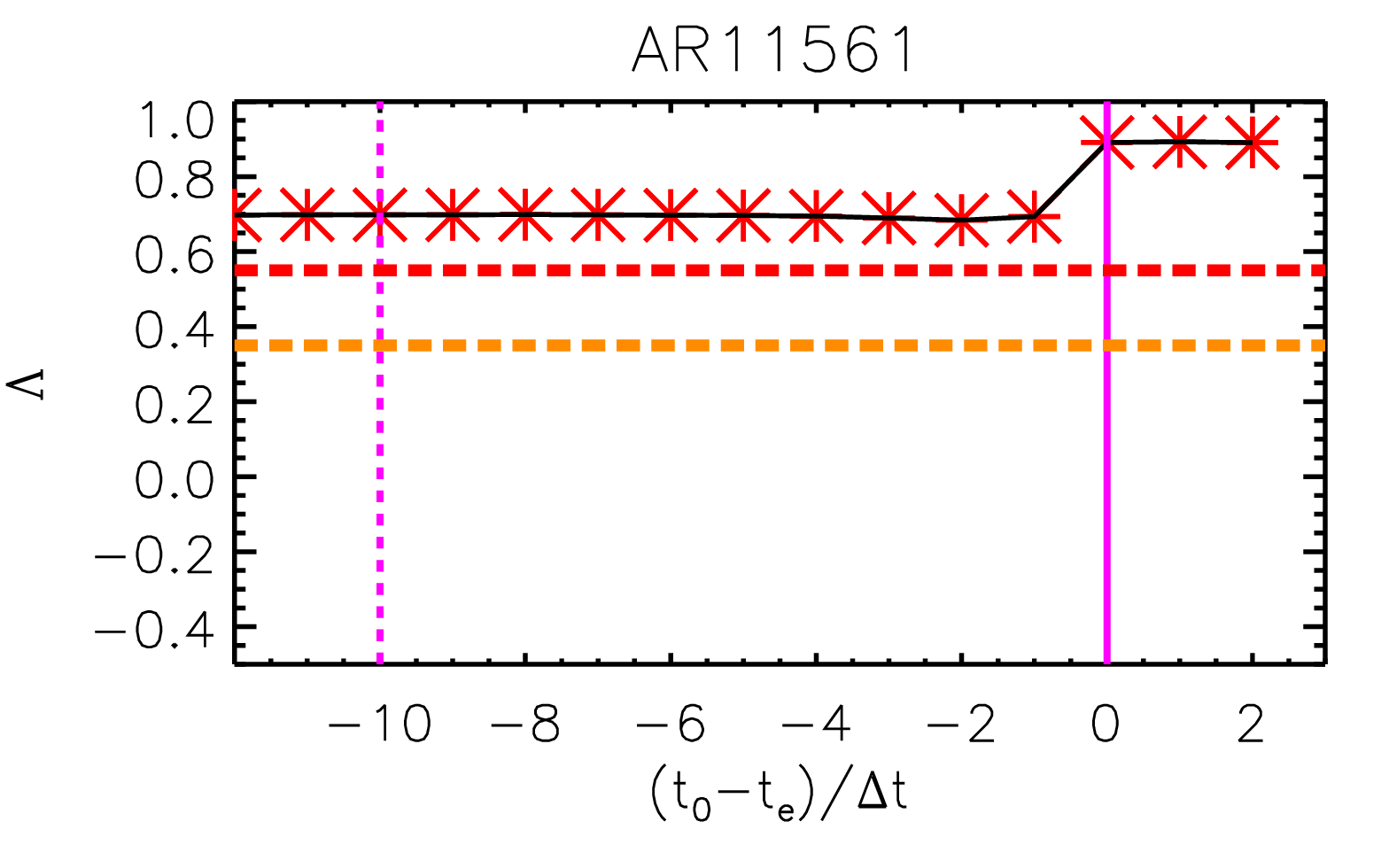}
\includegraphics[scale=0.28]{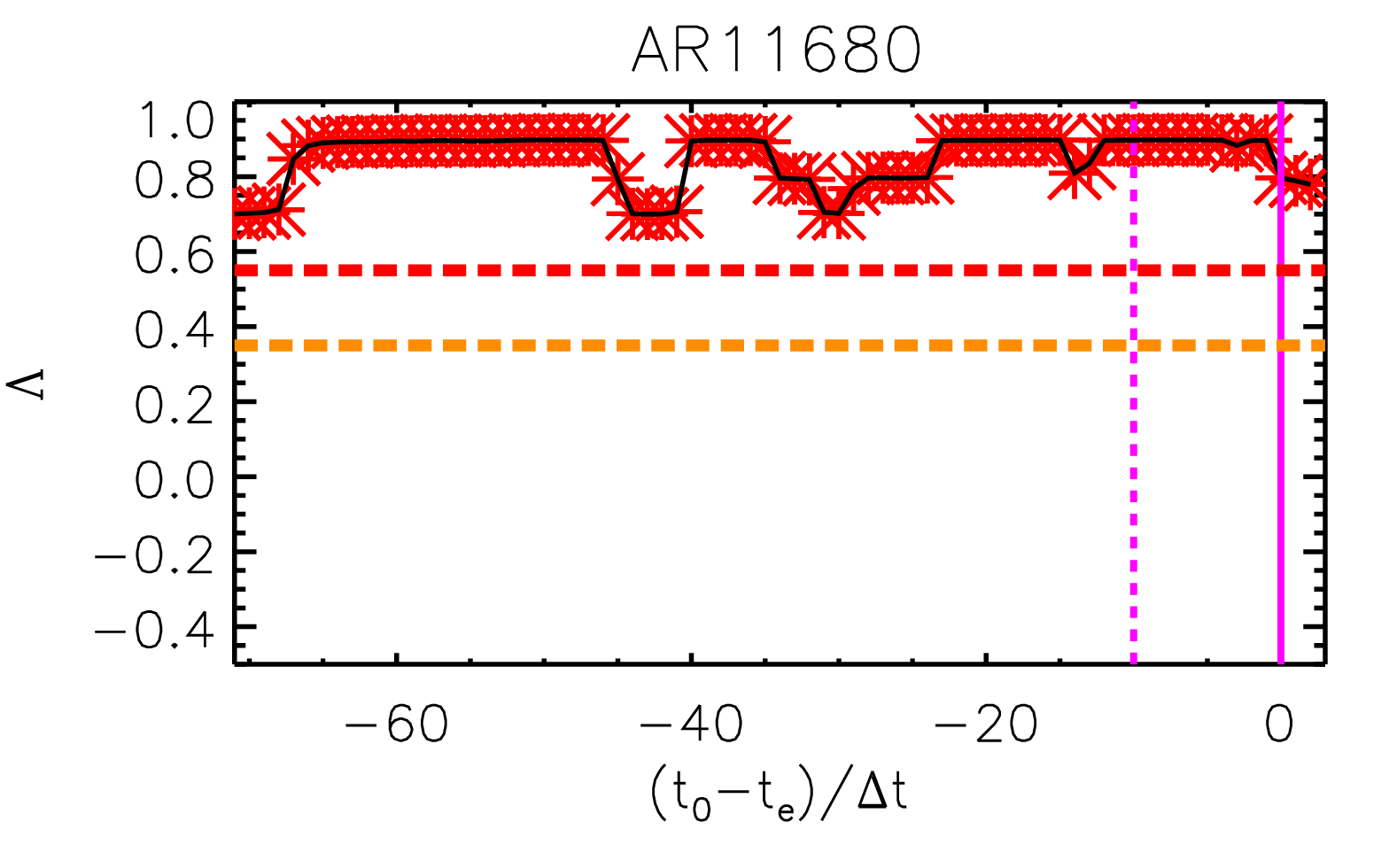}
\includegraphics[scale=0.28]{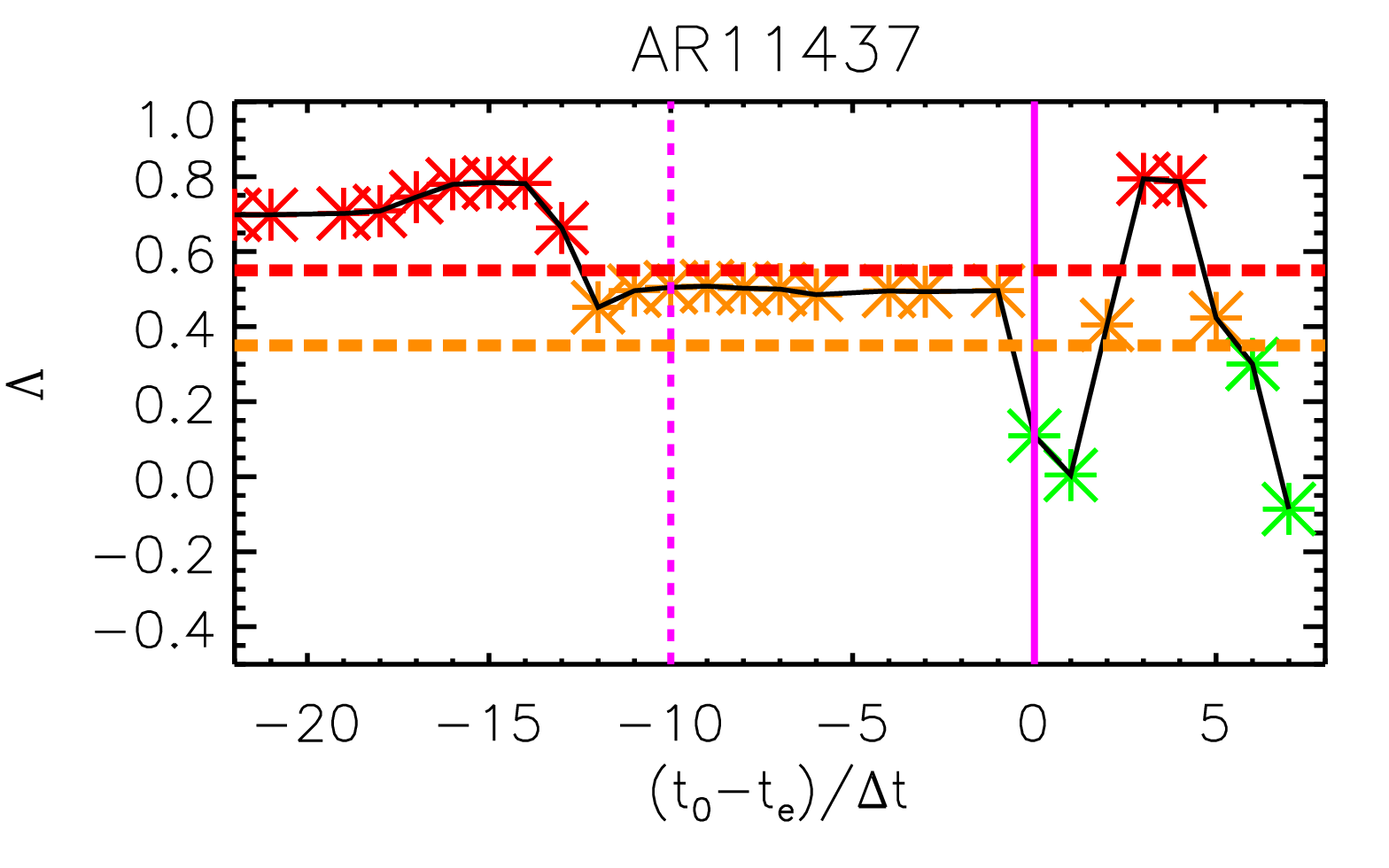}
\includegraphics[scale=0.28]{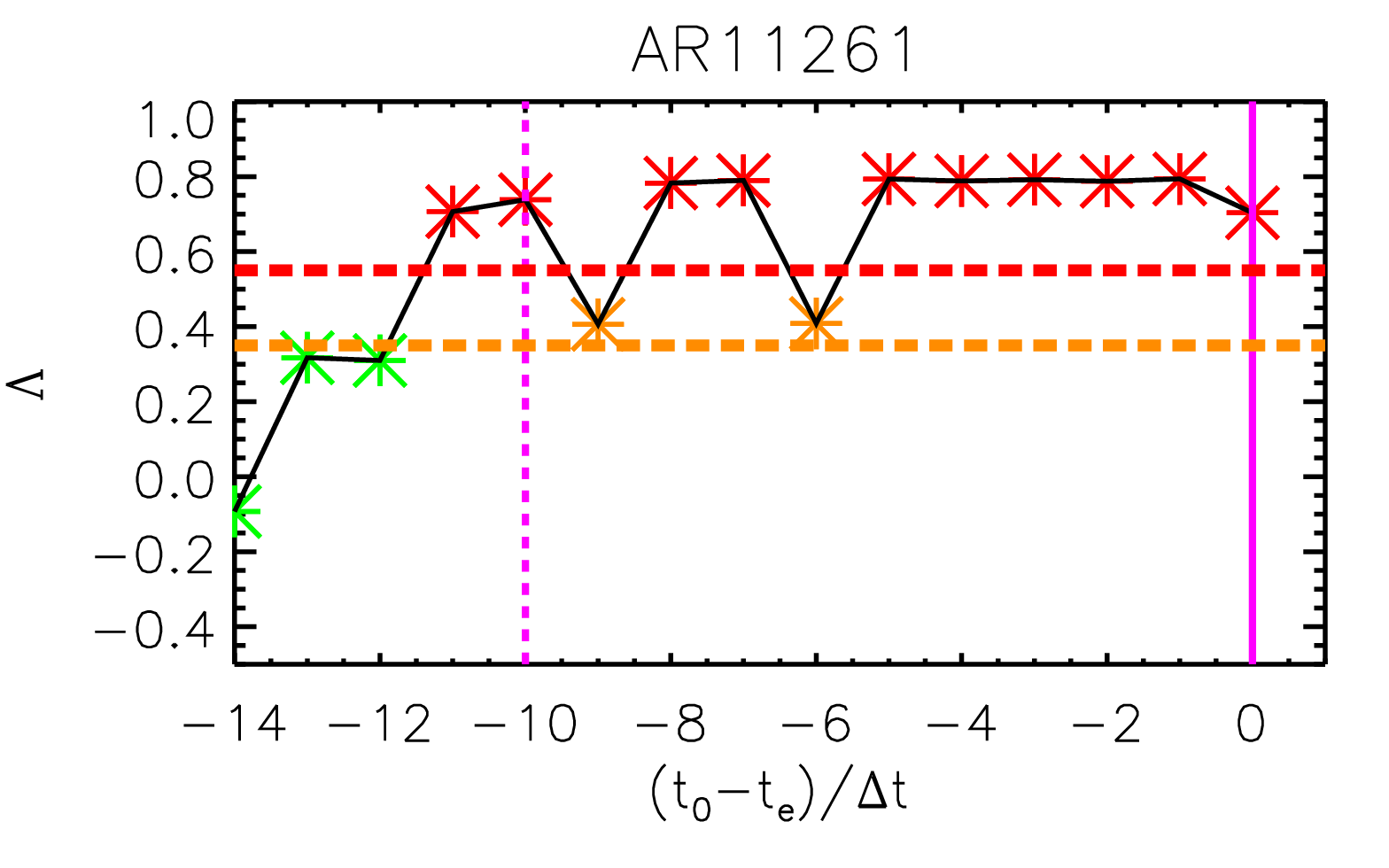}
\includegraphics[scale=0.28]{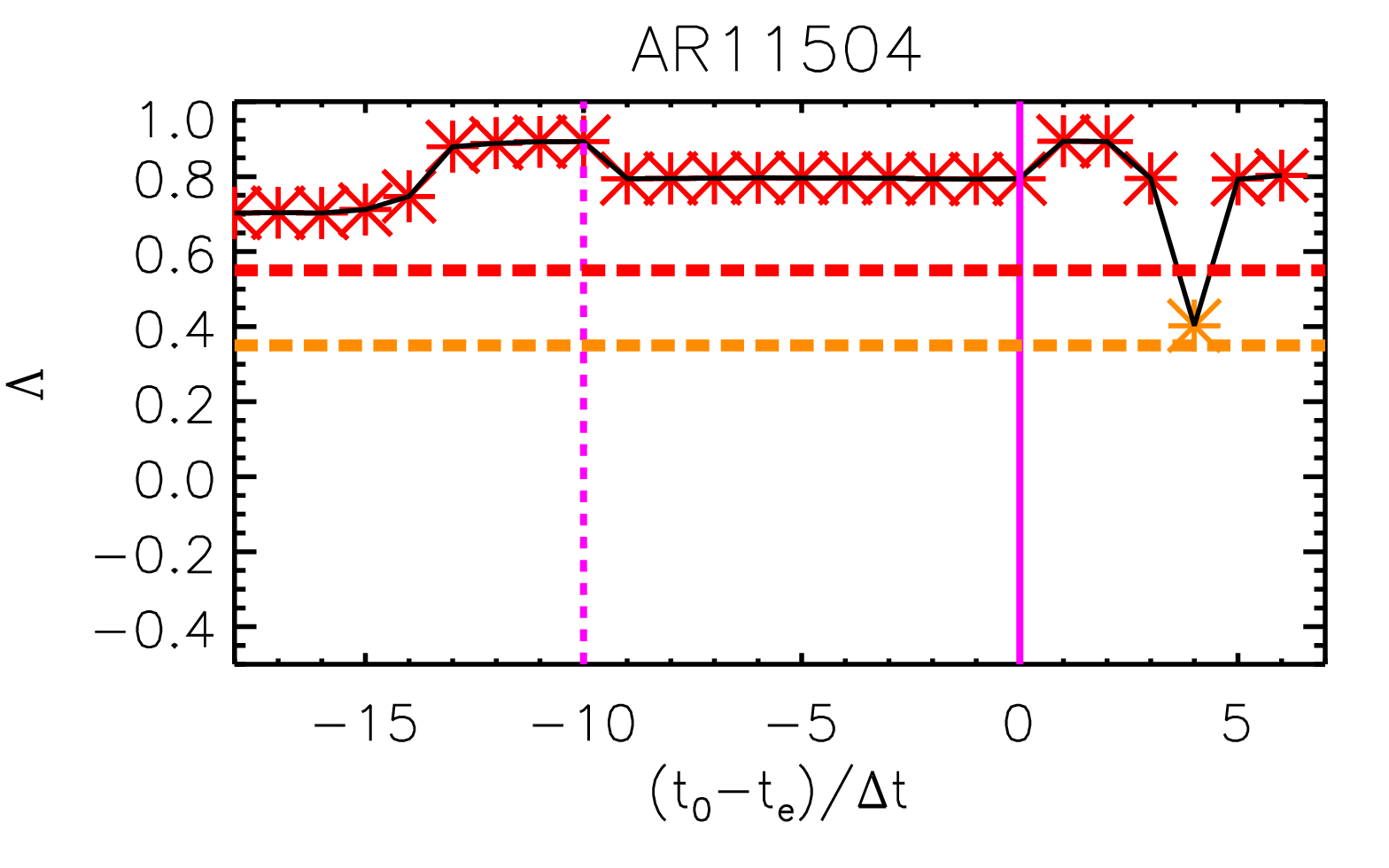}

\caption{The evolution of $\Lambda$ for the eruptive active regions as a function of $t_0$ (black line).
The asterisk symbols at each $t_0-t_e/\Delta t$ are the values of $\Lambda$ for each simulation with a given $t_0$.
Symbols are red for $\Lambda\geq 0.55$, amber for $0.35\geq\Lambda< 0.55$,
and green for $\Lambda< 0.35$.
The amber and red dashed horizontal lines represent these thresholds, respectively amber and red.
The magenta vertical lines represent $t_e$ (continuous) and $t-t_e=-10\Delta t$ (dashed).}
\label{lambda_eruptive}
\end{figure}
Fig.\ref{lambda_eruptive} shows the evolution 
of the value of $\Lambda$ (asterisks) as a function of 
$\left(t_0-t_e\right)/\Delta t$ for the active regions which were associated with 
observed eruptions. The asterisks are colour coded such that they are:
red when $\Lambda\geq 0.55$,
amber when $0.35\leq\Lambda< 0.55$,
and green when $\Lambda< 0.35$.
The solid magenta line represents the time of the observed eruption.
While the dashed magenta line gives the time
$t_0=t_f-10\Delta t$ where the first projected simulation includes the observed eruption 
within its projected time span.
These plots show only the time interval after the ramp-up phase of each simulation.

The ideal outcome
is that all of the asterisks in the interval between the dashed and continuous magenta lines
are red or amber, and that the asterisks are green outside of this interval.
However, this is not the case and so while this technique is useful 
in identifying eruptive active regions, it needs improvement to address 
the exact time window of the eruption within the active regions.
AR11561 and AR11680 show high values of $\Lambda$ at all times and are thus 
considered eruptive at all times. 
Similarly, AR11504 is considered eruptive for the vast majority of the time,
except near the end of the magnetogram time series where a decrease of $\alpha_{\Delta\zeta}$ reduces the value of $\Lambda$ after the time of the observed eruption.
AR11437 shows mixed results where the active region appears eruptive for $t_0<t_e-10\Delta t$ and then it decreases to an amber alert in the time frame before the
observed eruption.
However, it immediately decreases to a non-eruptive classification after the eruption,
with the exception of a couple of points.
AR11261 shows a rather successful classification, as it is initially categorised as non-eruptive
for $t_0<t_e-12\Delta t$ and then it increases to a red or amber alert during the 
time period when the active region is expected to be eruptive.
From these results it can be seen that in 4 out of the 5 active regions
$\Lambda$ successfully measures the risk of eruption in the active regions 
for the time periods leading up to the observed eruption. For the remain 
active region it would be borderline eruptive.

\begin{figure}
\centering
\includegraphics[scale=0.28]{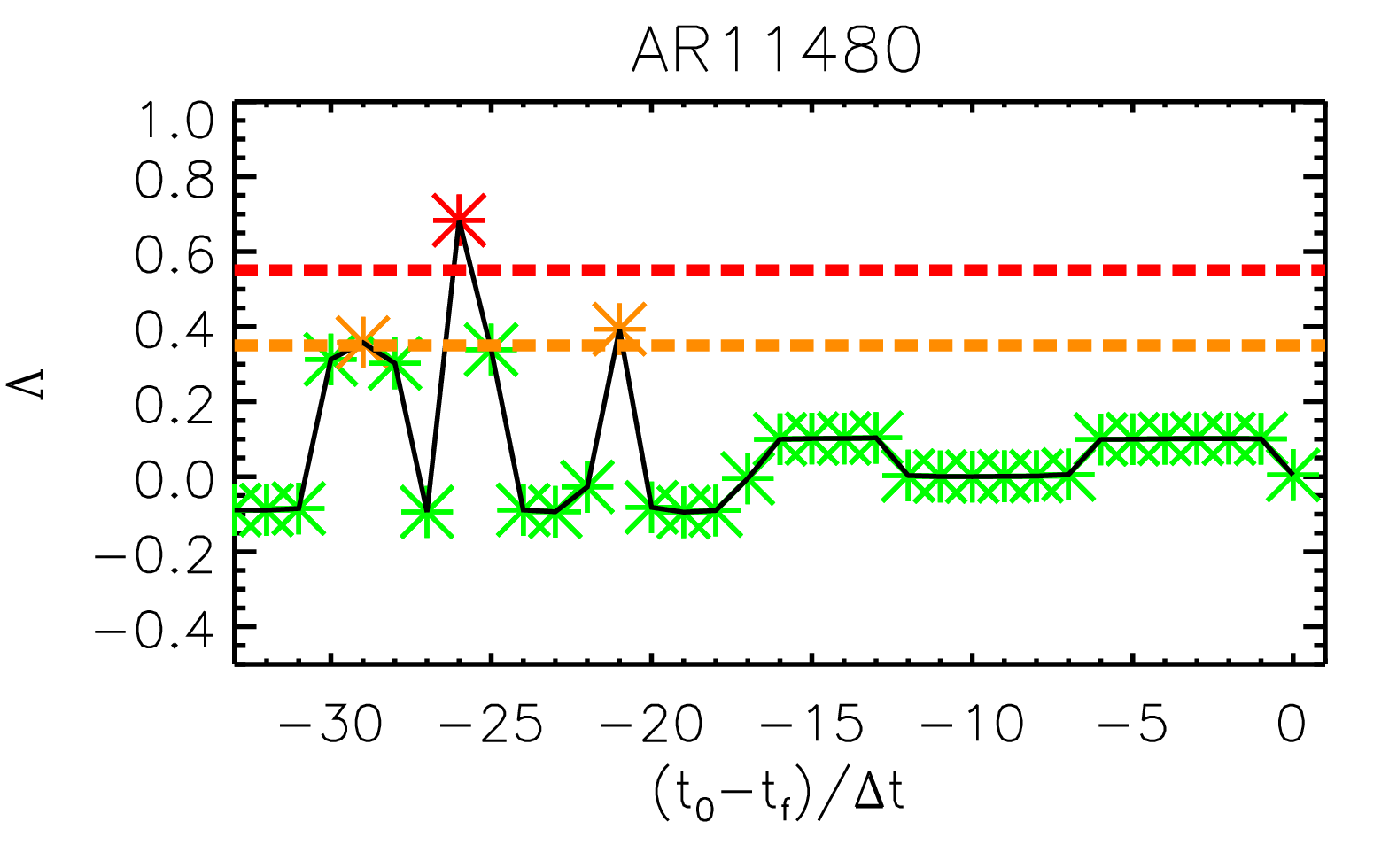}
\includegraphics[scale=0.28]{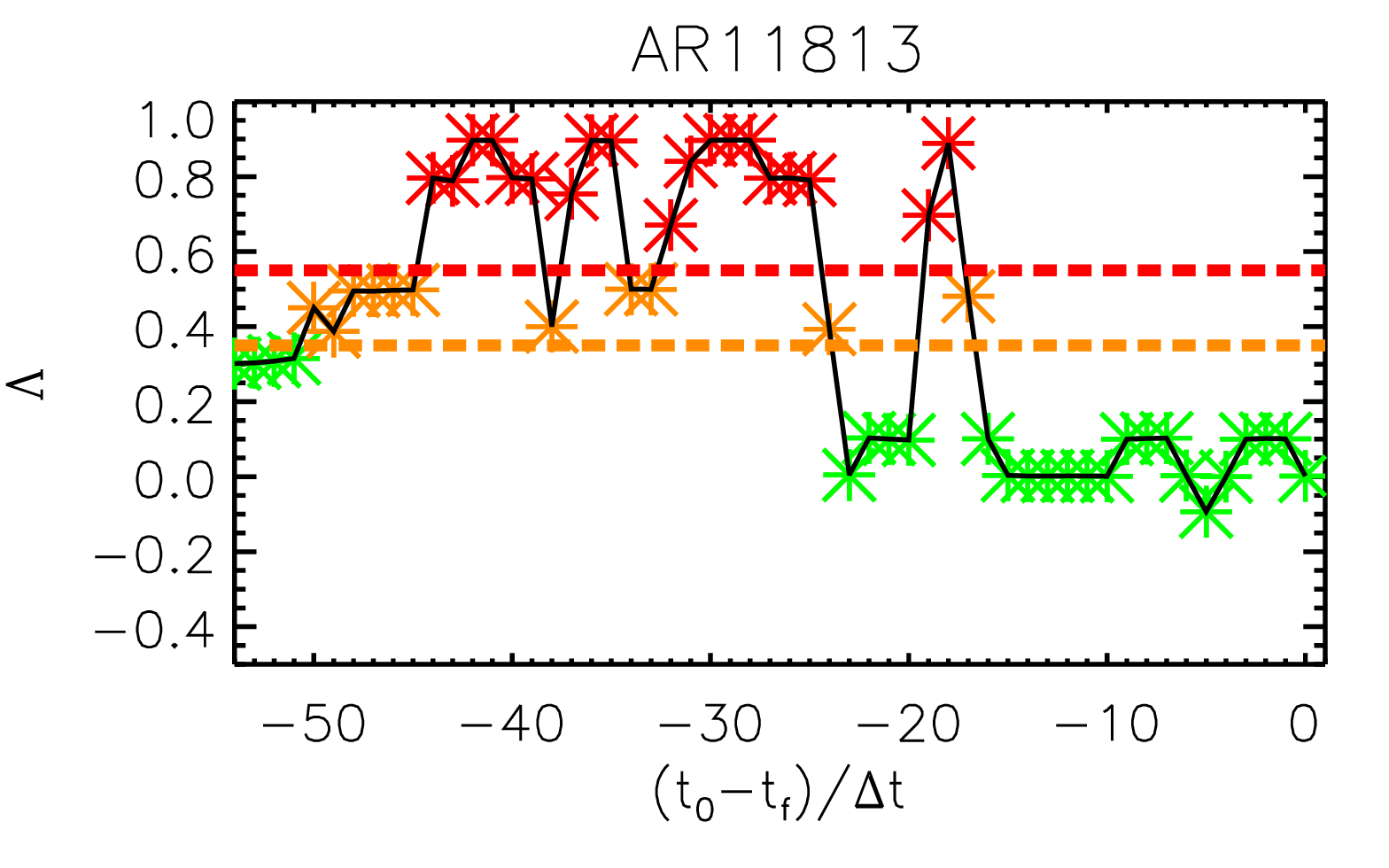}
\includegraphics[scale=0.28]{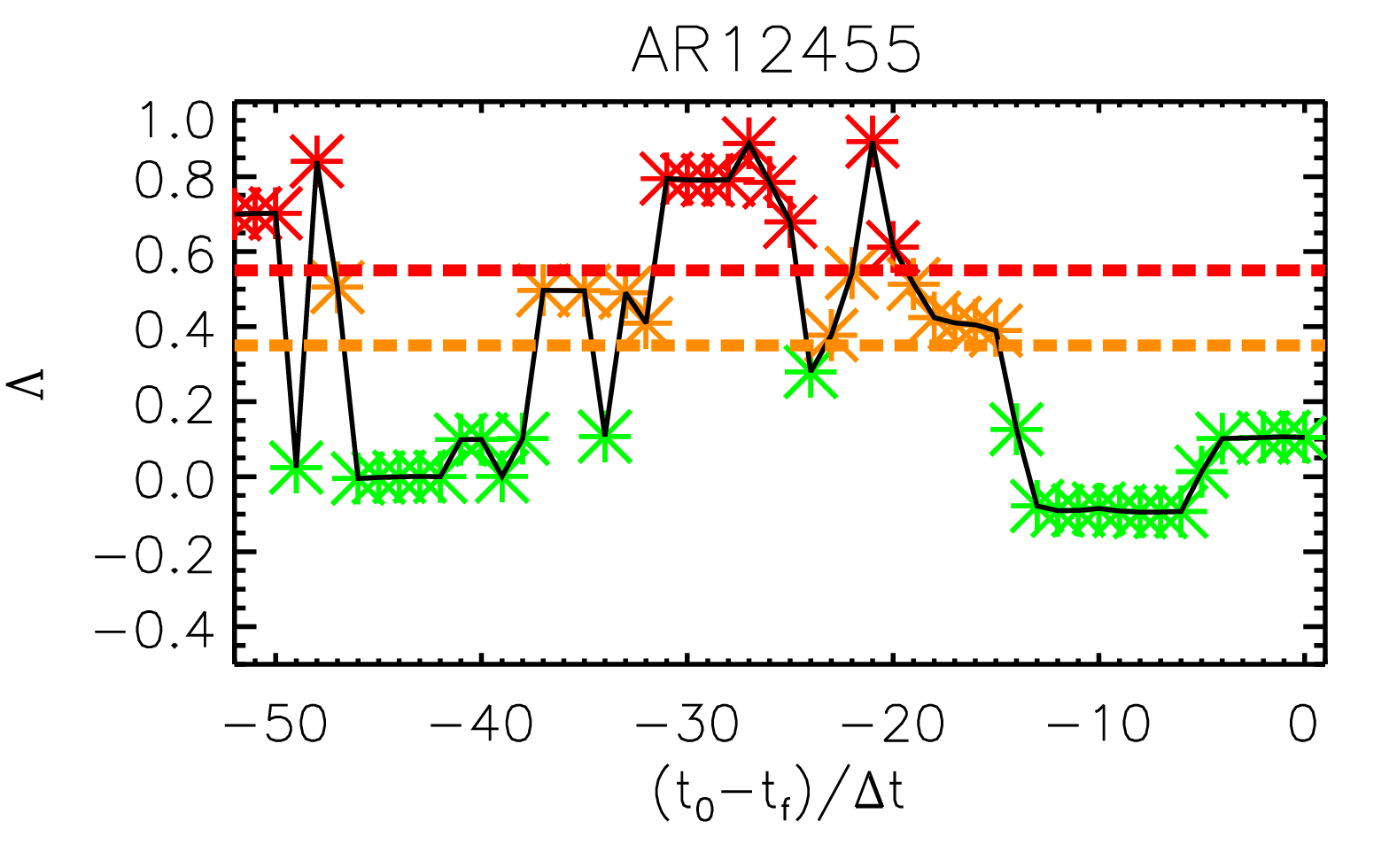}
\caption{
The evolution of $\Lambda$ for the non-eruptive active regions as a function of $t_0$ (black line).
The asterisk symbols at each $t_0-t_e/\Delta t$ are the values of $\Lambda$ for each simulation with a given $t_0$.
Symbols are red for $\Lambda\geq 0.55$, amber for $0.35\geq\Lambda< 0.55$,
and green for $\Lambda< 0.35$.
The amber and red dashed horizontal lines represent these thresholds, respectively amber and red.}
\label{lambda_noneruptive}
\end{figure}
Fig.\ref{lambda_noneruptive} shows the evolution of $\Lambda$
for the non-eruptive active regions, where we also find promising results.
A common feature of each plot is that the initial values of $\Lambda$, obtained
during the early projection simulations just after the ramp-up phase, are rather scattered. In contrast, as the simulations progress and more observational 
information is incorporated, the values of 
$\Lambda$ settle to values less than 0.35, where the active regions are correctly
classified as non-eruptive during these times.
In Paper I, we had already found a similar feature, 
where the magnetofrictional simulation description of the 3D magnetic field configuration becomes increasingly more accurate as we depart from the ramp-up phase.

\section{Conclusions}
\label{conclusions}

In this paper, we have extended the theoretical study of Paper I towards a 
possible practical application that aims to monitor active regions and 
identify which ones are more likely to produce an eruption. In particular, 
we use the same set of active regions as in Paper I (5 eruptive and 3 
non-eruptive) and simulate their 3D magnetic field evolution
with a data-driven magnetofrictional model. To obtain the future
risk of eruptions we develop a technique that uses a combination 
of observed and projected magnetograms as the lower boundary condition.
We then apply the metric, $\bar{\zeta}$, that we have introduced in 
Paper I to verify that it can discriminate between eruptive and 
non-eruptive active regions when using projected magnetograms over a 
fixed time period of around 10 magnetograms (10-16hrs).
The metric is based on three properties of the magnetic field that are 
linked to eruptions: the presence of magnetic flux ropes, the strength 
and direction of the Lorentz force and finally the Lorentz force heterogeneity. 
In this paper, we use the metric $\zeta\left(x,y,t\right)$
and quantities derived from it, such as
$\zeta_{max}\left(t\right)$ and $\bar{\zeta}$, to define a new parameter 
$\Lambda$ that, in this exploratory study is applied to the same set of
active region as in Paper I. We show that $\Lambda$
more clearly predicts the risk of an eruption 
in the next 10 to 16 hours (approximately 10 magnetograms at the 
cadence used in this work).

The parameter $\Lambda$ is constructed using a combination of 
quantities derived from $\zeta\left(x,y,t\right)$
and its value cannot exceed 1 by construction.
In our study we find values in the full range from 0-1
and we identify two empirical operational thresholds for $\Lambda$.
When $\Lambda\geq0.55$ there is a high risk that the active region will produce and eruption (red alert),
when $\Lambda<0.35$ the risk is instead rather low (green alert).
We also identify a medium risk when the value of $\Lambda$ is between the thresholds (amber alert). 
We find that in some 
cases we have clear indications of an eruption becoming more likely
as the value of $\Lambda$ exceeds the thresholds that we have empirically identified.
From the point of view of simple identification of eruptive active regions,
the technique clearly reaches its goal, as all five eruptive active regions are classified as eruptive for the majority of the simulated active region evolution.
Additionally, our technique classifies some of the eruptive active regions as 
non-eruptive when the evolution is still far from the eruption time or post-eruption, thus limiting false positives.
One limitation of the present technique is that it cannot yet identify the exact eruption time of a 
specific eruption. In the near future we hope to overcome this by improving
the derivation of the metrics $\zeta\left(x,y,t\right)$ and $\Lambda$ along with their practical applications. We note that this initial specification of the threshold may need to be revised when the sample of active regions considered is increased.

Presently, it remains to be determined whether a prediction of the eruption time is possible at all
with the spatial resolution of the current instruments.
The elusive nature of the eruption 
time can be a consequence of the sensitive nature of the mechanisms
that trigger solar eruptions. If it is so, 
it is natural that active regions remain in a metastable equilibrium for a long time before something at
smaller spatial scale or some external trigger causes the runaway from equilibrium and the eruption.
If this is the scenario we are facing, identifying eruptive active regions 
is the only achievable goal for the next few decades until new instruments
allow us to resolve the dynamics at smaller spatial scales. 

We also find that non-eruptive active regions are generally associated with low values of $\Lambda$
when $t_0$ is far enough from the end of the ramp-up phase,
providing a correct classification.
In contrast, a higher number of false positives are found near the end of the ramp-up phase in the simulation.
This phase corresponds to the 
time required for the simulations to depart from the initial 
potential magnetic configuration that we know to be an inaccurate
representation of the active region magnetic field.
We have already stressed in Paper I and in other works
how crucial it is to have a long time series of magnetograms
in order to reach a satisfactory detailed description of the
magnetic field in the active region.
Therefore, we believe that these uncertainties can be reduced 
if longer time series of magnetograms are available,
as we have shown in Paper I.
In Paper I, it was also shown that the projection simulations perform better in
reproducing the value of $\bar{\zeta}$ for the active regions
where we have a longer time series available.
To mitigate this problem, the best approach would be to develop the code so that the initial field condition is constructed from an NLFFF extrapolation based on a vector magnetogram. In the future, we intend to develop such a capability, which would allow extended testing on a wider sample of active regions, whose observations start days or weeks after they have emerged.

To put our work in context, 
the forecasting of solar eruptions is a topic that has attracted great 
interest in recent years \citep{2018ApJ...856....7H,2018ApJ...857..107L,2018JSWSC...8A..25L,2017ApJ...843..104L,2018ApJ...858..113N,2015ApJ...802L..21K,2017SoPh..292..159K,2018SoPh..293...96K}.
One of the most promising efforts has been made by the EU Flare Likelihood and Region Eruption Forecasting project \citep[FLARECAST;][]{2017AGUFMSA21C..07G}
that associate a number of features readily measurable from active region observations to the likelihood that an eruption will occur.
These studies have generally connected 
properties derived from observed magnetograms to some likelihood of eruption.
However, the 3D magnetic field
which is analysed in our paper inherently carries more
information about the stability of the coronal structures than 
photospheric observations alone. Existing NLFFF models, in 
particular the magnetofrictional model, allow computationally fast 
continuous simulations of the full 3D magnetic field, and 
the prediction obtained from this approach can be more detailed.
The initial work presented in Paper 1 and the present paper
provides a pathway for future research focused on 
developing a robust model to predict active region eruptions.
Here we have adopted the simplest approach in terms of the metric and 
projection technique, but we have nevertheless explored the possibility 
of identifying eruptive from non-eruptive active regions.

\begin{figure}
\centering

\includegraphics[scale=0.35]{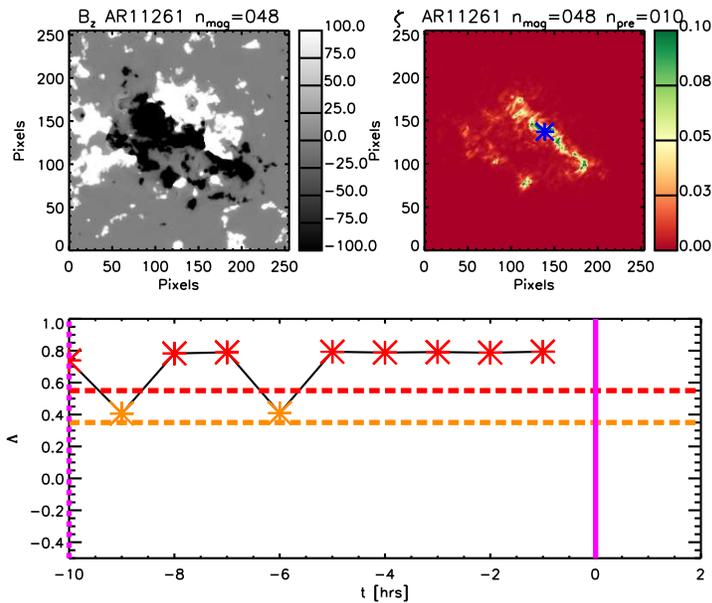}

\caption{Example of graphic interface for S$^2$WARM (movies available online).
The top panels are used to represent the most updated observed magnetogram (left)
and the projected distribution of $\zeta\left(x,y,t\right)$ (right), respectively.
The lower panel shows the evolution of $\Lambda$ at the time of the most recent magnetogram.
In this example for AR11261, we show the magenta lines that represent the time of the eruption (continuous line)
and the time of 10 magntograms before the eruption (dashed line), i.e. when 
the technique should start identifying the active region as erupive.}
\label{toolkit}
\end{figure}

For a more practical perspective, we propose that in the future that
this technique will be able to run automatically 
on space weather systems that focus on active regions monitoring.
We imagine that our framework, the St Andrews Space Weather Active Region Monitor (S$^2$WARM),
is particularly useful for issuing timely space weather warnings
or supporting scientific missions and their target selection for observation campaigns .
Fig.\ref{toolkit} and the corresponding movies available online show an example of how the technique
can be implemented in a visual interface,
where we simultaneously show the most recently available magnetogram of the active region at $t=t_0$ (upper left panel),  
the projected distribution of $\zeta\left(x,y,t\right)$ for $t>t_0$ (upper right panel)
and the evolution of $\Lambda$ (lower panel).
The panels are updated as new magnetogram observations and predictions are acquired.
This framework allows for a constant monitoring of the active regions present on the solar disk
and can facilitate decisions in space weather or mission control centres.
Naturally, it is expected that our framework is going to be benchmarked against 
already existing space weather predictive tools. For instance, forecasting tools such as MAG4 \citep{2011SpWea...9.4003F,2014AAS...22440204F} that is used by the National Oceanic and Atmospheric Administration (NOAA) and by NASA's Space Radiation Analysis Group (SRAG) and the FLARECAST \citep{2017AGUFMSA21C..07G,2018cosp...42E1181G}
platform that is developed by a consortium of European based institutions and includes 
some follow up applications using machine learning \citep{2018SoPh..293...28F}.
The direct comparison between these approaches can significantly empower our capacity
of responding to Space Weather hazards.

Despite the present technique not being currently able to predict 
the eruption time, it does open new opportunities to give more 
advanced warnings through the coupling of space weather models 
to the low solar corona. A number of space weather forecasting tools,
such as EUPHORIA \citep{2018JSWSC...8A..35P},
or WSA-ENLIL \citep{2000JGR...10510465A,2003AdSpR..32..497O}
rely on the cone model \citep{Xie2004}. The cone model
is a simplified description of the injection of CMEs into the solar wind.
The combination of the magnetofrictional relaxation technique along with the method of projected evolution
allows us to identify eruptive active regions and then run tentative space weather forecast simulations prior to the eruption with different speculated eruption times
to have broad predictions on the trajectory, speed, and properties of the Interplanetary CME (ICME).
Also, running successive magnetofrictional simulations,
as more observational data is acquired, will allow us to reduce the uncertainties in the predicted ICMEs properties.
In a more complex scenario, it is possible to couple the results of this approach with full MHD simulations
of the low solar corona as in \citet{Pagano2013a}, \citet{Rodkin2017}, and \citet{Pagano2018} to produce a more realistic insertion model of the CME into 
the solar wind.
This more realistic model could include the time evolution of the magnetic field vectors along with the mass and velocity of the ejected material.

Finally, it should be mentioned that this work is fundamentally based
on the analysis of a series of 3D magnetic field configurations 
of active regions. Starting from the previous works where the magnetofrictional model generates
an accurate representation of the 3D magnetic field configuration, we have then been guided by the physical interpretation of the results
to derive $\zeta\left(x,y,t\right)$, $\zeta_{max}\left(t\right)$, $\bar{\zeta}$,
and finally the prediction parameter $\Lambda$.
An alternative route that we have not explored yet is to use machine learning 
techniques in order to extract different proxies from the 3D configuration of the magnetic field that may outperform our current technique.
We have thus far preferred a physics-minded approach, as our goal is not only the practical application of this study for the benefit of space weather predictions and Solar Orbiter operations, but also a deeper understanding of the physics that govern solar eruptions.
\acknowledgments

This research
has received funding from the Science and Technology Facilities Council (UK)
through the consolidated grant ST/N000609/1 and the European Research Council
(ERC) under the European Union Horizon 2020 research and innovation
program (grant agreement No. 647214).
This work used the DiRAC@Durham facility managed by the Institute for
Computational Cosmology on behalf of the STFC DiRAC HPC Facility
(www.dirac.ac.uk). The equipment was funded by BEIS capital funding
via STFC capital grants ST/P002293/1, ST/R002371/1 and ST/S002502/1,
Durham University and STFC operations grant ST/R000832/1. DiRAC is
part of the National e-Infrastructure.
S.L.Y. would like to acknowledge STFC for support via the Consolidated Grant SMC1/YST025 and SMC1/YST037.
DHM would like to thank both the UK STFC and the ERC (Synergy Grant: WHOLE SUN, Grant Agreement No. 810218) for financial support.

%



\bibliography{ref}


\end{document}